\documentclass[preprint]{aastex}
\shorttitle{HST study of the stellar populations in NGC 2366}
\shortauthors{T. X. Thuan \& Y. I. Izotov}
\usepackage{natbib}

\begin{document}

\title{A {\sl HST} study of the stellar populations in the cometary 
dwarf irregular galaxy NGC 2366{\footnote{Based on observations 
obtained with the NASA/ESA {\it{Hubble Space Telescope}} through the Space 
Telescope Science Institute, which is operated by AURA,Inc. under NASA
contract NAS5-26555.}}}

\author{Trinh X. Thuan}
\affil{Astronomy Department, University of Virginia,
    Charlottesville, VA 22903}
\email{txt@virginia.edu}

\and

\author{Yuri I. Izotov}
\affil{Main Astronomical Observatory, National Academy of Sciences of Ukraine, 03680, Kyiv, Ukraine}
\email{izotov@mao.kiev.ua}

\begin{abstract}

We present $V$ and $I$ photometry of the resolved stars in the cometary dwarf 
irregular galaxy
NGC 2366, using Wide Field Planetary Camera 2 images obtained with
the {\sl Hubble Space Telescope}. The resulting
color-magnitude diagram reaches down to $I$ $\sim$ 26.0 mag. It reveals
not only a young population of blue main-sequence stars (age $\la$ 30 Myr)
but also an intermediate-age population of blue and red
supergiants (20 Myr $\la$ age $\la$ 100 Myr), and an older evolved populations 
of asymptotic giant branch (AGB) stars (age $\ga$ 100 Myr) and red giant branch
(RGB) stars (age $\ga$ 1 Gyr). 
The measured magnitude $I$ = 23.65 $\pm$ 0.10
mag of the RGB tip results in a distance modulus $m-M$ = 
27.67 $\pm$ 0.10, which corresponds to a distance of 3.42 $\pm$ 0.15 Mpc, in
agreement with previous distance determinations.
The youngest stars are associated with the bright complex of H {\sc ii} 
regions NGC 2363 $\equiv$ Mrk 71 in the southwest extremity of the galaxy. 
As a consequence of the diffusion and relaxation 
processes of stellar ensembles, the older the stellar population is, the 
smoother and more extended is its spatial distribution. An underlying
population of older stars is found throughout the body of NGC 2366. The
most notable feature of this older population is the presence of numerous 
relatively
bright AGB stars. The number ratio of AGB to RGB stars 
and the average absolute 
brightness of AGB stars in NGC 2366 are appreciably higher than in the 
BCD VII Zw 403, indicating a younger age of the AGB stars in NGC 2366. 
In addition to the present
burst of age $\la$ 100 Myr, there has been strong star formation activity 
in the past of NGC 2366, from $\sim$ 100 Myr to $\la$ 3 Gyr ago. 

\end{abstract}


\section{Introduction}

In the hierarchical model of galaxy formation, large galaxies result from
the merging of smaller structures. These building-block dwarf galaxies are
too faint and too small to be studied directly at large redshifts. However,
we stand a much better chance of understanding these with local examples. The
dwarf irregular galaxy NGC 2366 is suitable 
for carrying out a detailed study with
the {\sl Hubble Space Telescope} ({\sl HST}). A member of the M81 group of
galaxies \citep{Ka85}, it is at a distance of 3.44 Mpc, as 
determined by
Cepheid variables \citep{To95}. Using the brightest stars as distance
indicators, \citet{Ti91} also find a distance of 
3.4 Mpc\footnote{At this distance, 1\arcsec\ corresponds to a linear size of
16 pc.}. Thus, NGC 2366 is close enough for the galaxy to be resolved into 
stars by {\sl HST} and color-magnitude diagrams (CMDs) of its stellar
populations can be constructed to determine its star formation history. 
NGC 2366 is the prototype of the interesting class of ``cometary''
galaxies described by \citet{Lo86}. It is characterized by an 
elongated body -- the tail of the ``comet'' -- and a bright starburst region
at its southwest end -- the head of the ``comet''.

   The complex of bright high-excitation H {\sc ii} regions in the
southwest part of NGC 2366 is called NGC 2363 
$\equiv$ Mrk 71. Several 
abundance measurements have been made for NGC 2363 
\citep{Pe86,Ma91,Go94,Iz97,No00}. 
They yield an oxygen
abundance 12 + log O/H $\sim$ 7.9, or 1/10 solar. Modeling of the physical
conditions in NGC 2363, based on observations from the literature, has been 
carried out by \citet{Lu99}.
Using Fabry-Perot and optical spectroscopic observations, \citet{Ro91,Ro92}
have discovered an expanding supernova bubble in NGC 2363.
   Besides NGC 2363, there are other H {\sc ii} regions in NGC 2366. 
\citet{Ro96} have derived the abundances for several of these H {\sc ii} regions.
They find their oxygen abundances 12 + log O/H to be in the range 8.1 -- 8.3,
slightly higher than that of Mrk 71.

Several observations of NGC 2366 have been made in the 21 cm line 
\citep{Hu81,Th81,We86,Hu01,T04}. All these determinations give
similar neutral hydrogen gas masses,
$M$(H {\sc i}) = 8.0 $\times$ 10$^8$ $M_\odot$\footnote{\citet{Hu01}
by misreading a decimal point mistakenly noted that the value of \citet{Th81}
was 10 times higher than theirs. In fact, the two values are
in good agreement.}, so that the ratio
of integrated H {\sc i} content to blue luminosity 
$M$(H {\sc i})/$L(B)$ is 1.36 $M_\odot$/$L(B)_\odot$.
A striking feature of the H {\sc i} intensity map of NGC 2366 is
the presence of two parallel ridges about 6\arcmin\ (5.8 kpc) long
running along the major axis of the optical body and separated by a zone
nearly devoid of H {\sc i}. \citet{Hu01} and \citet{T04} have interpreted 
those structures to be a H {\sc i} ring inclined with respect to the line
of sight by an angle of $\sim$ 60\arcdeg. The H {\sc i} ring is surrounded
by diffuse neutral gas emission, forming an extended envelope around the
optical body.

   \citet{Dr00} have used {\sl HST}
and ground-based observations to obtain CMDs of NGC 2363 and discuss 
its star formation history and that of neighboring star clusters. 
However, their analysis was mostly
restricted to young stellar populations. Of note is the discovery of a rare
luminous blue variable (LBV) 
star in NGC 2366 by \citet{Dr97}, 5\arcsec\ east of NGC 2363, and 
presently the brightest optical source in the galaxy. 
The LBV star has been erupting since 1994. 
\citet{Dr01} have monitored the
photometric and spectroscopic time variations of the LBV
star with the {\sl HST} and have constructed a physical model for it.

The general properties and evolutionary status of NGC 2366 have been discussed
by \citet{No00}. They
have used photometric and spectroscopic ground-based observations
to put constraints on the age of the older stellar populations in NGC 2366. 
\citet{No00} concluded that the light of NGC 2366 is
dominated by stellar populations with age not exceeding 3 Gyr, significantly
lower than the typical age of $\ga$ 5 Gyr derived for the underlying
stellar component of other dwarf galaxies.

However their conclusions are based on population synthesis of
spectroscopic observations,
the {\sl HST} observations of \citet{Dr00,Dr01} not being deep
enough for studying old stellar populations. To remedy the situation, 
we have obtained deep {\sl HST} Wide Field and Planetary Camera 2 (WFPC2) $V$ 
and $I$ images of NGC 2366. We use these data to look anew at both young and
old stellar populations of the cometary dwarf galaxy and discuss its 
evolutionary status. 
We describe the observations in Sect. 2. The distance to NGC 2366 is 
derived in Sect. 3. The stellar populations and star formation history
of NGC 2366 are discussed in Sect. 4. 
 We summarize our findings in Sect. 5.

\section{Observations}

\subsection{Data reduction}

  We have obtained {\sl HST} images of NGC 2366 on 2000 December 12 during 
cycle 9 with the WFPC2 through filters F555W and F814W, which we will refer 
to hereafter as $V$ and $I$.
The observations were broken into five subexposures in the $V$ filter and 
into three subexposures in the $I$
filter to permit identification and removal of cosmic rays. 
The total exposure time was 6700s in $V$ and 4100s in $I$. The galaxy 
was positioned
so that the brightest H {\sc ii} region NGC 2363 is located in the PC frame,
to take advantage from its twice as good spatial resolution. 
The WFPC2 was oriented in order the major axis of the galaxy lies along the
diagonals of the PC and WF3 frames. The scale of the WFPC2
is 0\farcs046 per pixel in the PC frame and 0\farcs102 per pixel in the 
WF frames. Because the considerable major axis of NGC 2366 
($a_{I_{25}}$ $\sim$ 300\arcsec) 
is larger than the combined diagonal lengths of the PC and WF3 frames
($\sim$ 160\arcsec), the 
northermost parts of the galaxy could not be imaged.

   Preliminary processing of the raw images including corrections for
flat-fielding was done at the Space Telescope Science Institute through the 
standard pipeline. Subsequent reductions were carried out at the Main
Astronomical Observatory of the Ukrainian Academy of Sciences
and the University of Virginia
using IRAF{\footnote{IRAF is the Image Reduction and Analysis Facility
distributed by the National Optical Astronomy Observatory, which is
operated by the Association of Universities for Research in Astronomy
(AURA) under cooperative agreement with the National Science Foundation
(NSF).}} and STSDAS{\footnote{STSDAS: the Space Telescope Science Data 
Analysis System.}}.
 Cosmic rays were removed and the images in each filter were combined using 
the CRREJ routine. We found that all subexposures in a given filter 
coregistered to better than $\sim$ 0.2 pixels.

\begin{figure}
\figurenum{1}
\epsscale{0.9}
\plotone{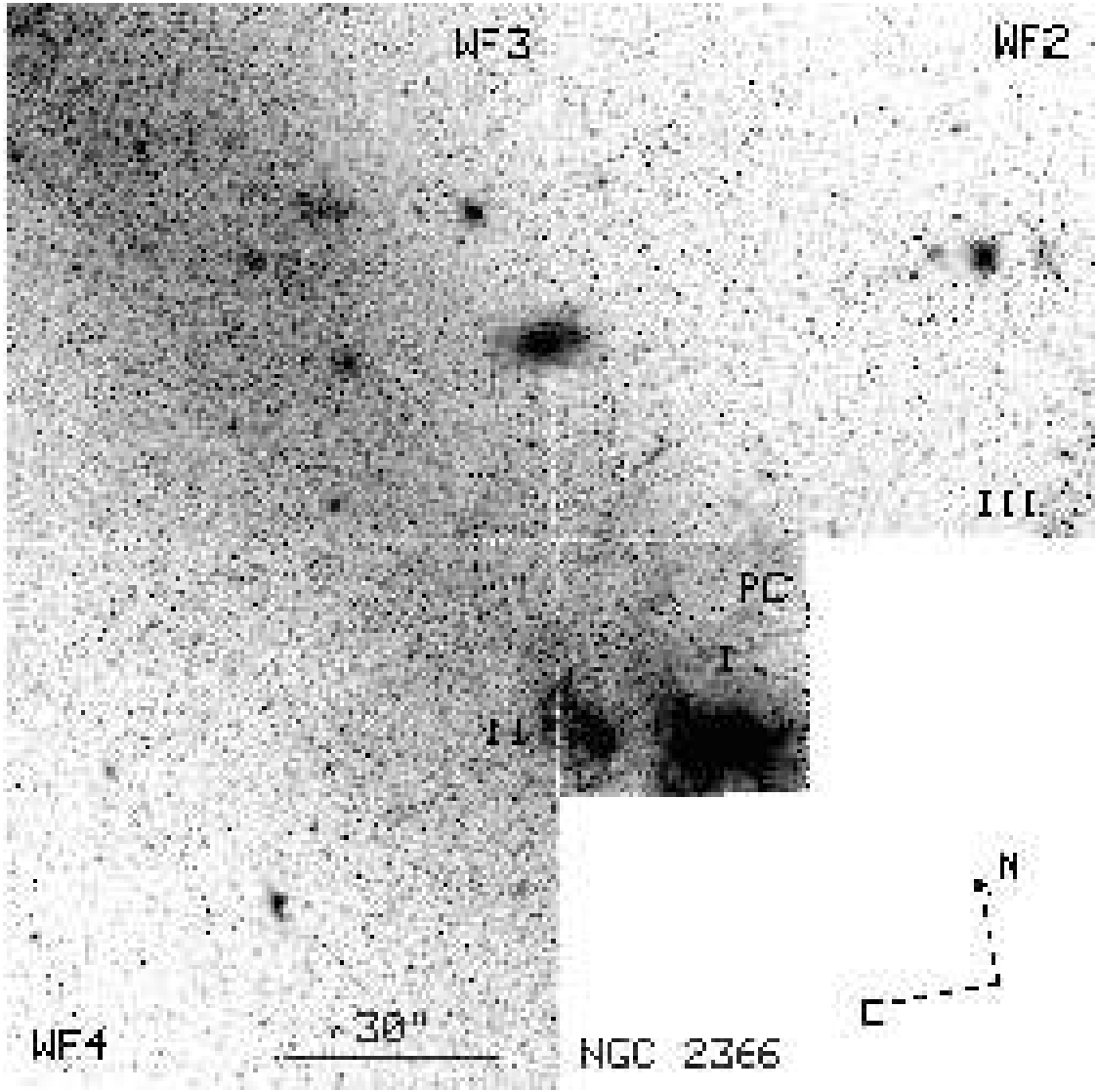}
\caption{{\sl HST} $V$ mosaic image of NGC 2366. Regions I 
($\equiv$ NGC 2363 $\equiv$ Mrk 71), II and III are labeled following the 
notation of \citet{Dr00}.
The regions of present star formation reside nearly all in the PC 
frame. 
\label{Fig1}}
\end{figure}

Fig. \ref{Fig1} shows the mosaic $V$ image of NGC 2366. 
We shall follow the notation of \citet{Dr00} in labeling the three main 
H {\sc ii} regions. The two brightest ones, I and II, are on the PC chip,
while we see a small part of III in the bottom right corner of the WF2 frame.
The extended low-surface-brightness body of the galaxy is visible in all WF
and PC frames.
Several background galaxies can be seen in the field of NGC 2366, 
the brightest of them being the spiral galaxy to the north of region II.

\subsection{Construction of color-magnitude diagrams}

   The superior spatial resolution of the {\sl HST}/WFPC2 images combined 
with the proximity of NGC 2366 permits to resolve the galaxy into
individual stars and study its
stellar populations by means of CMDs. We used the DAOPHOT 
package in IRAF for point-spread-function (PSF)-fitting photometry. 
The PSFs are derived separately for each of the PC
and WF fields and in each filter using the brightest isolated stars in 
each image.
The background level in the PC field was measured 
in an annulus with radii 4 and 6 pixels (0\farcs18 and 
0\farcs28) around each source and 
subtracted. 
The sky level in the WF fields was measured in 
an annulus with radii 3 and 5 pixels (0\farcs30 and 
0\farcs51). 
The PSF-fitting 
stellar photometry was done by adopting the zero points of \citet{Ho95b} 
and a 2-pixel fitting radius for the PC frames and a 1.5-pixel fitting radius 
for the WF frames.
The detectability limit is set to the 3$\sigma$ threshold above the local
background noise. To convert 
instrumental magnitudes with an aperture
radius of 2 or 1.5 pixels to the magnitudes corresponding to 
the \citet{Ho95a} calibrating aperture radius of
0\farcs5 (respectively 11 and 5 pixels for the PC and WF frames), 
we need to derive aperture corrections. For this, we compared 
PSF-fitted magnitudes of bright isolated 
stars with the magnitudes of the same stars
measured with the aperture photometry technique within an 0\farcs5 aperture.
For the PC frames we obtained the corrections $V_{\rm ap}$(0\farcs5) -- 
$V_{\rm fit}$ = --0.63 mag and $I_{\rm ap}$(0\farcs5) -- $I_{\rm fit}$ 
= --0.77 mag. For the WF frames the corresponding corrections are --0.39 mag
in both $V$ and $I$. Our derived corrections are very similar to 
those obtained by \citet{Ho95a}.

\begin{figure}
\figurenum{2}
\epsscale{0.8}
\plotone{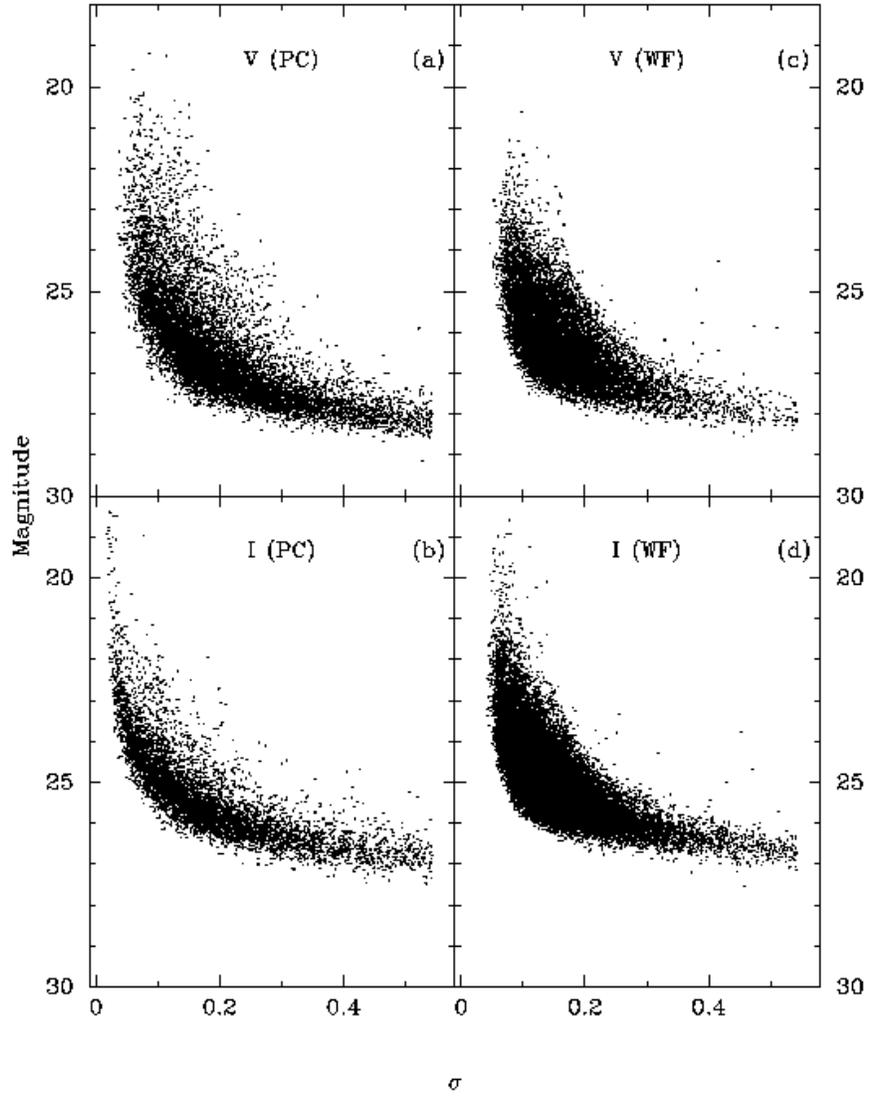}
\caption{Photometric error $\sigma$ as a function of stellar apparent
magnitude in both $V$ and $I$ images in the PC frame (panels a and b),
and in the WF2,WF3,WF4 frames (panels c and d). The $V$ photometry is
$\sim$ 1 mag deeper than the $I$ photometry and goes to a limiting
magnitude of $\sim$ 28.5 mag. \label{Fig2}}
\end{figure}

Stars with a sharpness out of the --1.0 -- +1.0
range in both PC and WF frames were eliminated
to minimize the number of false detections.
Correction for charge-transfer efficiency loss has been
done according to the prescriptions of \citet{Do02}.
Figure \ref{Fig2} shows the distribution of photometric errors as a 
function of $V$ and $I$ magnitudes as determined by DAOPHOT.
It is seen that errors are about 0.2 mag at $V$ = 27 mag and 
$I$ = 26 mag for both PC and WF frames. 
They increase to about 0.4 mag at $V$ = 28 mag and 
$I$ = 27 mag. 

The total numbers of recovered stars in both PC and WF
frames are respectively 51632, 33564 and 22121 in 
the $V$ band, the $I$ band and in both bands,
adopting a matching radius of 1 pixel. The corresponding numbers of recovered 
stars in the PC frame only are 9888, 5918 and 3415. That more than a third 
of the stars are not matched is due to the combination of incompleteness
effects and an increasing number of false detections at faint magnitudes.
The transformation of instrumental magnitudes to the Johnson-Cousins $UBVRI$ 
photometric system as defined by \citet{La92} was performed according to the 
prescriptions of \citet{Ho95b}. The magnitudes and colors of
point sources were corrected for Galactic interstellar extinction adopting 
$A_V$ = 0.12 mag and $A_I$ = 0.07 mag \citep{S98}.

We have carried out a completeness analysis for each of the frames
using the DAOPHOT routine 
ADDSTAR. For each magnitude bin listed in Table \ref{Tab1}, 
we have added artificial 
stars amounting to $\sim$ 5\% of the total number of real stars detected in 
each frame.
We then performed a new photometric reduction 
using the same procedure as the one applied to the original frame, and checked 
how many added stars were recovered in this magnitude bin. 
This operation was repeated 10
times for each frame and for each magnitude bin and the results were averaged. 
The completeness factor in each magnitude bin 
defined as the percentage of recovered artificial stars is shown 
in Table 1.
The completeness limit of the PC image is acceptable in the 25 -- 26 mag
range, $\sim$ 73\% and $\sim$ 66\% respectively in $V$ and $I$,
but drops to $\sim$ 30\% and $\sim$ 2\% 
in the 27 -- 28 mag range. The completeness limits of the WF2 and WF4
images are comparable to that of the PC image. However, because of the larger 
crowding, the completeness limits of the WF3 image are worse,
being $\sim$ 50\% in $V$ and $\sim$ 30\% in $I$ in the 25 -- 26 mag range. 

   Fig. \ref{Fig3} shows the $I$ vs $V-I$ CMD
of NGC 2366 derived from all frames. It can be seen that NGC 2366
contains diverse stellar populations in a variety of evolutionary stages:
main-sequence (MS), blue loop (BL), red supergiant (RSG), asymptotic giant 
branch (AGB), and red giant branch (RGB) stars.

\section{Distance determination}

\begin{figure}
\figurenum{3}
\epsscale{0.8}
\plotone{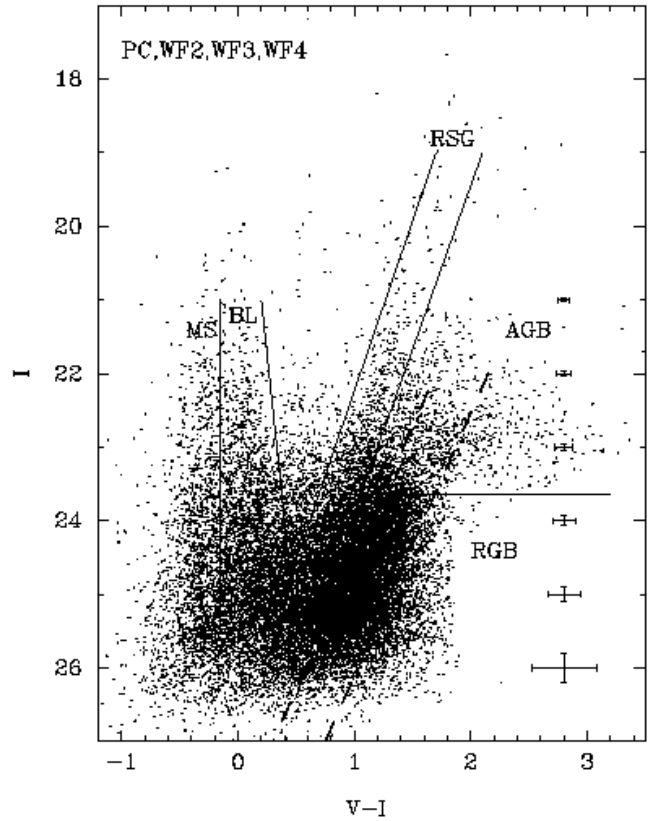}
\caption{The $V-I$ vs $I$ color-magnitude diagram of NGC 2366 from all frames.
The photometric errors are shown by horizontal bars at several magnitude 
levels. The solid lines delimit the regions occupied by stars in various
evolutionary stages: main-sequence (MS), blue loop (BL), red supergiant (RSG),
asymptotic giant branch (AGB) and red giant branch (RGB) stars. The dashed 
lines delimit the stars that are used to construct Fig. \ref{Fig4}.
\label{Fig3}}
\end{figure}

\begin{figure}
\figurenum{4}
\epsscale{0.7}
\plotone{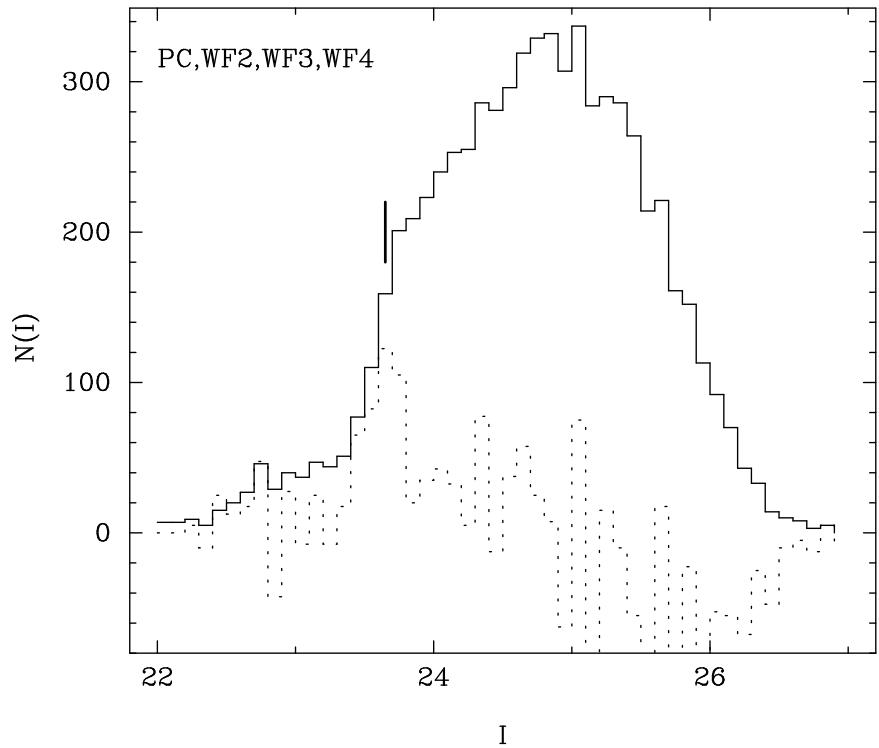}
\caption{The distribution of the red giant branch stars as a function
of the $I$ apparent magnitude (solid line) in all frames (PC, WF2, WF3, WF4).
The dotted line shows the derivative of the RGB star number distribution.
The location of the RGB tip is derived to be at 23.65 mag and is marked by a
vertical line. \label{Fig4}}
\end{figure}

The detection of RGB stars allows to derive the distance to NGC 2366 using
the observed magnitude $I$ of the tip of RGB stars (TRGB). This
technique is based on the observed constancy of the
absolute magnitude $M_{I}$ $\approx$ --4.05 mag of TRGB stars
in old globular stellar clusters 
\citep{Da90}. In the CMD, the TRGB is signaled by a sharp 
drop in the number of RGB stars toward brighter $I$ magnitudes.
This drop can be seen distinctly in the CMD of Fig. \ref{Fig3}. 
To quantify the drop, we have plotted the number distribution of the RGB 
stars in steps of 0.1 mag in Fig. \ref{Fig4} (solid line).
We have considered those RGB stars located in the CMD (Fig. \ref{Fig3})
to the blue of the straight line defined by  the pair of points 
($I$,$V-I$) = [(27,0.75);(22,2.15)] (right dashed line) 
to minimize the contribution of the AGB stars, and to the red 
of the straight line defined by 
($I$,$V-I$) = [(27,0.3);(22,1.7)] (left dashed line) to minimize
the contribution of the more massive red helium burning stars.
 The dotted line in Fig. \ref{Fig4} shows the 
derivative of the distribution. The location of the TRGB is determined by the
first large increase in both the RGB luminosity function and its derivative
and is marked by a vertical tick mark in Fig. \ref{Fig4}.
We obtain $I$(TRGB) = 23.65 $\pm$ 0.10 mag.

The distance modulus is derived from the equation 
$m-M$ = $I$(TRGB) -- $M_{I}$(TRGB). The absolute $I$ magnitude is 
defined as $M_{I}$(TRGB) = $M_{\rm bol}$(TRGB) -- $BC$($I$), where 
$M_{\rm bol}$(TRGB) is the bolometric magnitude of the TRGB and $BC$($I$)
is the bolometric correction to the $I$ magnitude.
The latter quantity is given as a function of the
$V-I$ color of the TRGB by $BC$($I$) = 0.881 -- 0.243$(V-I)_{\rm TRGB}$,
while the bolometric magnitude depends on metallicity and is given by 
$M_{\rm bol}$(TRGB) = --0.19[Fe/H] -- 3.81 \citep{Da90,Le93}. Hence 
for distance determination, knowledge
of the stellar metallicity is necessary. Usually [Fe/H] is derived 
using the $V-I$ color of RGB stars at the absolute magnitude $M_{I}$ = --3.
As pointed out by \citet{Le93} from analysis
of Yale theoretical isochrones, this is because $V-I$ is mainly
dependent on metallicity and not very much on age. 

\begin{figure}
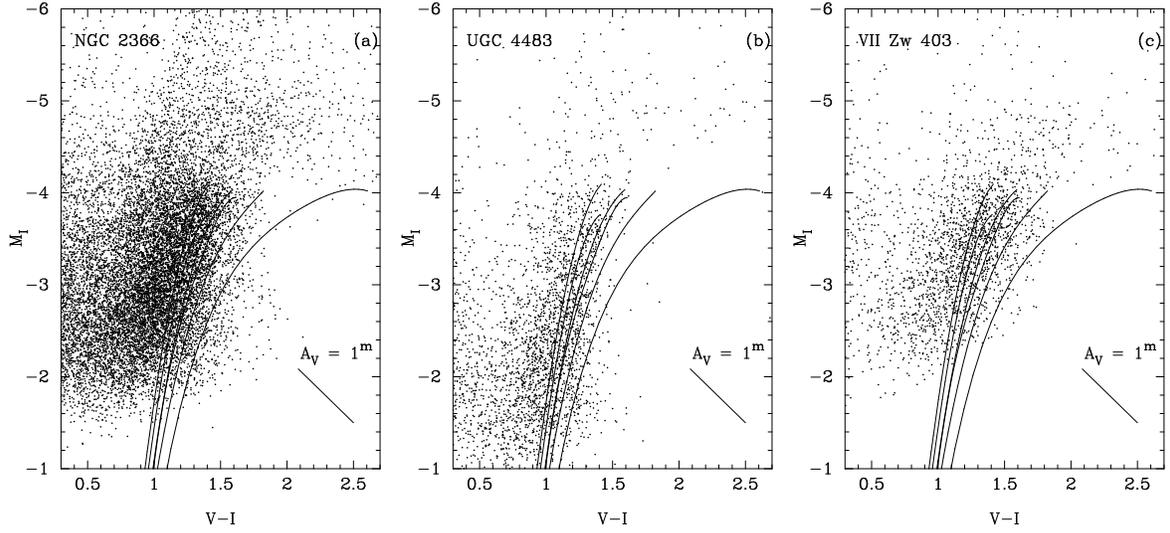

\figurenum{5}
\epsscale{0.3}
\plotone{f5a.ps}
\plotone{f5b.ps}
\plotone{f5c.ps}
\caption{The absolute magnitude $M_{I}$ vs $V-I$ color diagram of
the RGB stars in (a) NGC 2366, (b) the BCD UGC 4483 and c) the BCD 
VII Zw 403. 
The adopted distance moduli are respectively $m-M$ = 27.67, 27.63 
and 28.25 for NGC 2366, UGC 4483 and VII Zw 403.
The magnitudes and colors are corrected for extinction with $A_V$ = 0.12
mag, 0.11 mag  and 0.15 mag respectively. Solid lines
are isochrones for, from left to right, the globular clusters M15 
([Fe/H]=--2.17), NGC 6397 (--1.91), M2 (--1.58), NGC 6752 (--1.54), NGC 1851
(--1.29) and 47 Tuc (--0.71) \citep{Da90}.
Note the numerous AGB stars in NGC 2366 which are $\sim$ 0.3 -- 0.4 mag
brighter than those in VII Zw 403 \citep{Ly98,Sc98}, 
but are comparable in brightness to those in UGC 4483 \citep{Iz02}.
\label{Fig5}}
\end{figure}

In Fig. \ref{Fig5}a we show the RGB region of the CMD as
obtained from all frames, adopting a distance modulus $m-M$ = 27.67 mag
(to be discussed in the next paragraph).
We show by solid lines and from left to right the isochrones for the Galactic
globular clusters M15, NGC 6397, M2, NGC 6752, NGC 1851 and 47 Tuc 
with respective metallicities [Fe/H] = --2.17, --1.91, --1.58,
--1.54, --1.29 and --0.71 (Da Costa \& Armandroff 1990). 
Figs. \ref{Fig5}b and \ref{Fig5}c will be discussed in Section 4.4.
   It can be seen from Fig. \ref{Fig5}a that the clump of red stars
in the NGC 2366 CMD is broad with a significant part located to the
blue of the lowest metallicity isochrone, that of the globular cluster M15. 
We believe that the relatively large width of the red star distribution is 
real and that it is not 
due to uncertainties in the photometry, as these are 
small (Fig. \ref{Fig2}). 
The broad distribution implies that 
star formation in NGC 2366 has occurred for a relatively long period. 
The clump consists of a mixture of red helium burning stars and RGB
stars of different ages.

We set the $V-I$ color of the TRGB to be equal to the color of the reddest
RGB stars, ($V-I$)$_{\rm TRGB}$ $\sim$ 1.5. These stars 
are presumably the oldest ones in NGC 2363 and hence they are the most 
appropriate for comparison with globular cluster isochrones.
These reddest stars are fitted reasonably well by the isochrone
of the globular cluster M2 with [Fe/H] = --1.58.
For this metallicity,
$M_{I}$(TRGB) = --4.02 mag. The resulting distance modulus of NGC 2366 
is then $m-M$ = 
27.67 $\pm$ 0.10. The corresponding distance to the galaxy is
$D$ = 3.42 $\pm$ 0.15 Mpc, in excellent agreement with the values $D$ = 3.4 Mpc
derived by \citet{Ti91} from the brightest stars and $D$ = 3.44 Mpc by 
\citet{To95} from Cepheid variables. It is
larger than the value of 2.9 Mpc obtained by \citet{Ap95}, also
from the brightest stars. Because $M_I$(TRGB) is only slightly dependent on
the $V-I$ color, uncertainties in $(V-I)$(TRGB) have little influence on the
derived distance.

The above distance determination is based on the assumption that the reddest 
RGB stars in NGC 2366 have the age of globular cluster stars, i.e. are 
$\sim$ 10 Gyr old. 
For younger ages of the RGB stars,
theoretical isochrones, e.g. those of the Padua group \citep{Be94,Gi96,Gi00}
or the Geneva group \citep{Le01}, may be more appropriate.
These are most often used for the analysis
of resolved stellar populations in Local Group and relatively nearby galaxies.
However, as discussed by several authors \citep[e.g.,][]{Da90,Ly98,Iz02},
these theoretical isochrones do have problems: they do 
not fit those observed for globular clusters. In particular, 
\citet{Iz02} have shown that, except for the isochrones with 
$Z$ = 0.001 from \citet{Be94},
all 10 Gyr Padua isochrones 
do not reproduce the observed ones for 
globular clusters, being too blue
and not extending to bright enough absolute magnitudes. 
Furthermore, the isochrones that do fit were obtained from stellar evolutionary
models calculated with older opacities. Isochrones obtained from
models with new opacities for the same metallicity are again too blue
\citep{Gi00}. As for the theoretical 10 Gyr isochrones based on 
the Geneva stellar 
evolutionary models, they are too red compared to those 
observed for globular clusters. However, despite all these uncertainties,
Padua models \citep{Be94} can provide a good guide to the dependence of
$M_I$(TRGB) on the age of the RGB stars, since in this case we are considering
relative rather than absolute values. These models 
predict that $M_I$(TRGB) is nearly constant for RGB stars with 
ages $\ga$ 3 Gyr.
This implies that the distance of 3.42 Mpc derived above for NGC 2366 is
valid as long as its oldest RGB stars are $\sim$ 3 Gyr or older, 
which is likely the case for NGC 2366 (see section 4.4).


\section{Stellar populations}

   Several generations of stellar populations are evidently present in the 
CMD of NGC 2366 (Fig. \ref{Fig3}),
suggesting that star formation in this galaxy has occurred 
during the past several Gyr.
Ongoing star formation is evidenced by the presence
of the supergiant H {\sc ii} regions I and II 
at the southwestern edge of the galaxy.
Past star formation is inferred from the presence of MS stars 
with ages of only a few Myr, of BL and RSG stars with ages ranging
from $\sim$ 20 Myr to $\sim$ 100 Myr, of AGB stars with ages 
$\ga$ 100 Myr and of RGB stars with ages $\ga$ 1 Gyr.

\begin{figure}
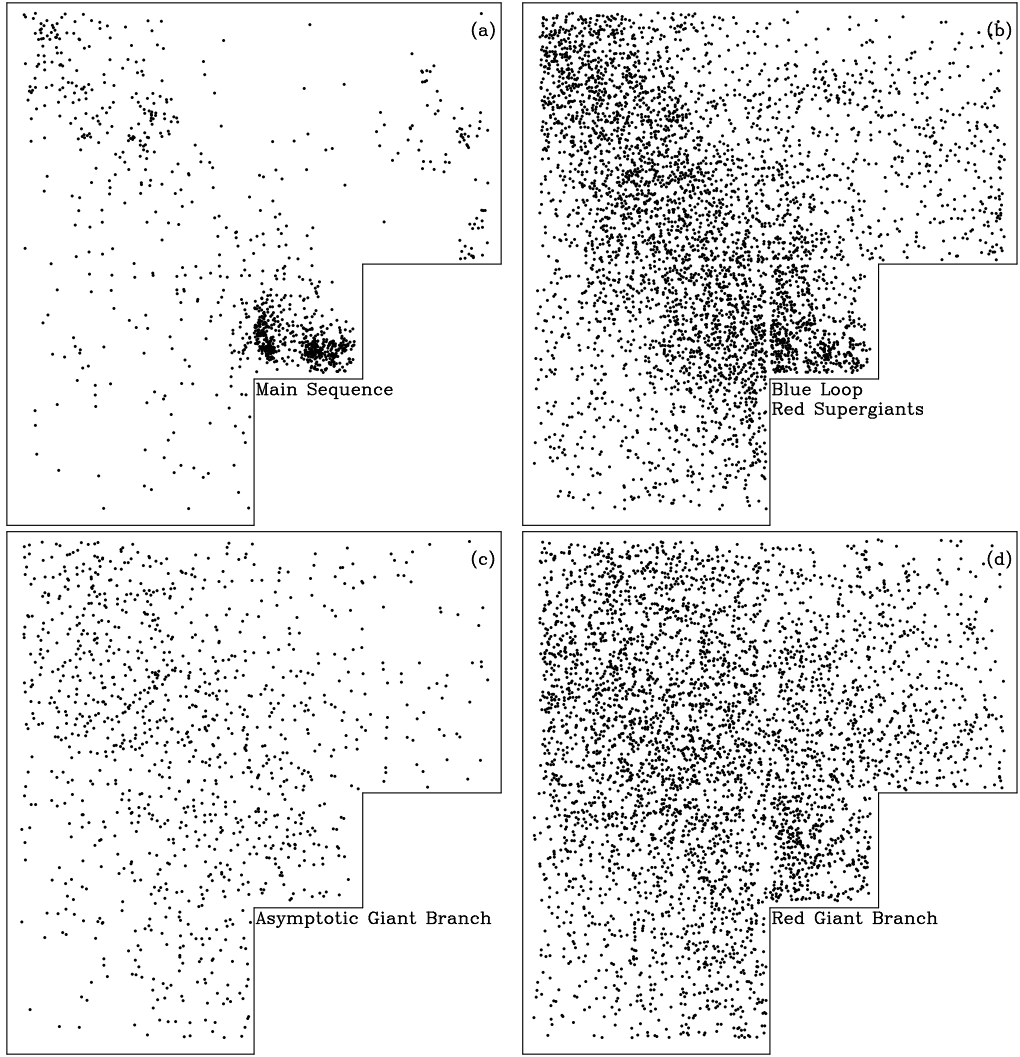

\figurenum{6}
\epsscale{0.4}
\plotone{f6a.ps}
\plotone{f6b.ps}
\plotone{f6c.ps}
\plotone{f6d.ps}
\caption{The spatial distribution of the (a) main-sequence, (b) blue loop and 
red supergiant,
(c) asymptotic giant branch and (d) red giant branch
stars. Stars with progressively larger ages from (a) 
to (d) are distributed over a wider and wider area and more and more smoothly.
The orientation is the same as in Fig. \ref{Fig1}. 
\label{Fig6}}
\end{figure}

   The spatial distribution of these different stellar populations,
as delimited in Fig. \ref{Fig3}, is shown in
Fig. \ref{Fig6}. To minimize incompleteness effects 
(Table \ref{Tab1}) and photometric errors (Fig. \ref{Fig2}) which become
important for $I$ $\geq$ 25 mag, we will consider  
in Fig. \ref{Fig6} and in all the following discussions on star 
counts only sources that are brighter than $I$ = 25 mag. 
It can be seen that a large number of the MS stars 
(Fig. \ref{Fig6}a)
are located in the H {\sc ii} regions I, II, and III as well as in those in the
northeast part of the elongated main body. They show a clumpy distribution,
reflecting the compactness of the H {\sc ii} regions. 
But there are also some MS
stars that are more evenly distributed over the whole body of the galaxy,
out to the edges of the WFPC2 frames.
BL and RSG stars are distributed more smoothly than MS stars
(Fig. \ref{Fig6}b), however their surface density is larger in the 
high-surface-brightness part of the main body. As for the
AGB and RGB stars, they show a smooth distribution over the whole WFPC2 
field of view (Fig. \ref{Fig6}c -- \ref{Fig6}d). There is a clear
correlation of the spatial extent of a stellar population with its age: 
the older the stellar population, the smoother and more extended is its 
spatial distribution.
Such population gradients have been known for a long time, 
starting with the pioneering studies of the Milky Way 
by Walter Baade in the 1950's. 
They have also been observed in other dwarf galaxies resolved by
{\sl HST} \citep[e.g.,][]{Ly98,Sc98,Cr00,Iz02}.
They are 
likely a consequence of the diffusion and relaxation processes of stellar
ensembles. We discuss next selected regions of NGC 2366 in more detail.

\subsection{Regions I and II}

\begin{figure}
\figurenum{7}
\epsscale{0.45}
\plotone{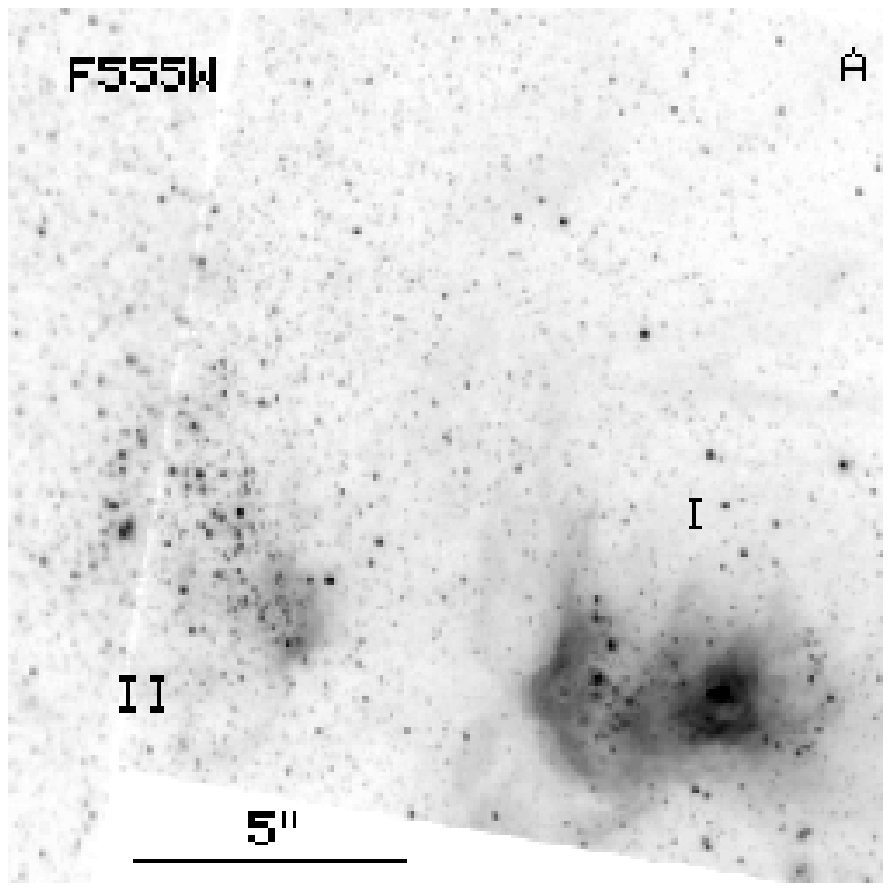}
\plotone{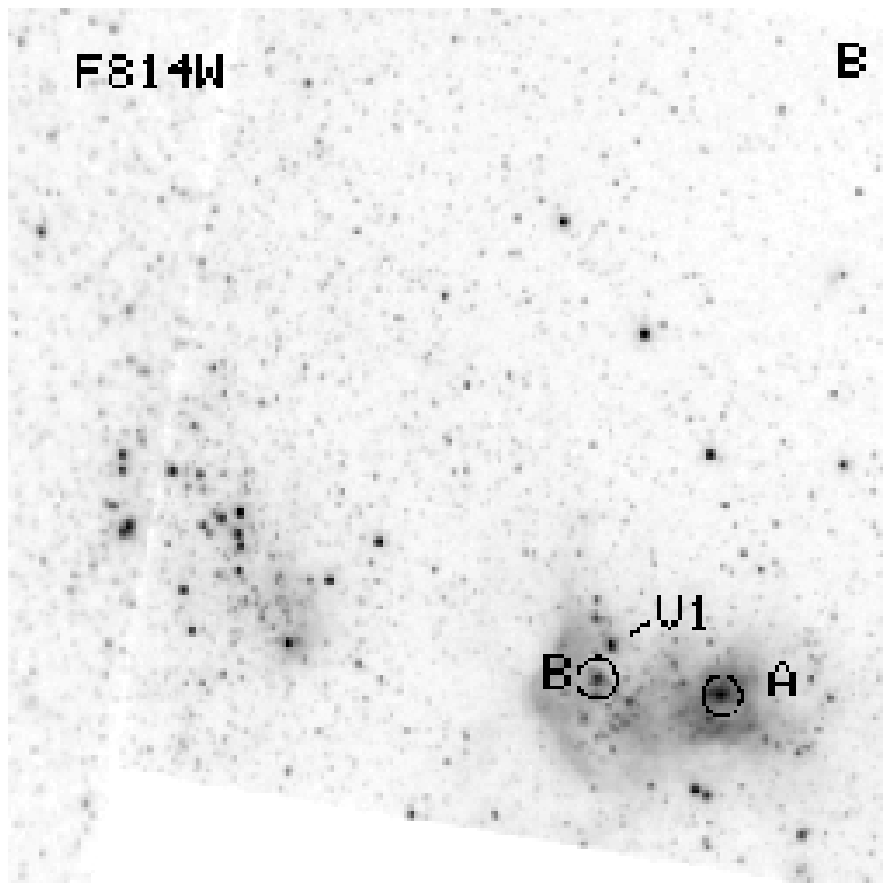}
\caption{PC $V$ (a) and $I$ (b) mosaic images of regions I and II shown 
with a logarithmic brightness scale.
North is up and east is to the left. The young compact ionizing clusters 
A and B and the luminous blue variable star V1 \citep{Dr00} are labeled in 
(b). \label{Fig7}}
\end{figure}

Most of the regions I ($\equiv$ NGC 2363 $\equiv$ Mrk 71) and II have been 
imaged with the PC, although some parts of region II are also in the WF4 frame
(Fig. \ref{Fig1}). Fig. \ref{Fig7} shows zoomed $V$ and $I$ 
views of the two regions. Widespread ionized gas emission resulting mainly from
strong [O {\sc iii}] $\lambda$5007 line emission, can be seen
in the $V$ image (Fig. \ref{Fig7}a). However, extended
ionized gas emission is also present in the $I$ image (Fig. \ref{Fig7}b),
in this case being mainly gaseous continuum emission. Region I contains
two young compact clusters which we label A and B following the notation
of \citet{Dr00}.
They are marked by circles in Fig. \ref{Fig7}b. The 
LBV star discovered by \citet{Dr97} is labeled as V1.
Unfortunately, the LBV star is saturated in both our $V$ and $I$ images
and cluster A is saturated in the $V$ image, preventing us from performing 
photometry of these two objects in those bands. 
Fig \ref{Fig7}a shows that the ionization of the gas in region I is
mostly caused by cluster A.
The ionization of the gas in region II is not as important, suggesting that
extremely young massive stars are absent there. Fig. \ref{Fig7}b shows also
the presence of numerous bright RSG stars in region II.

\begin{figure}
\figurenum{8}
\epsscale{0.45}
\plotone{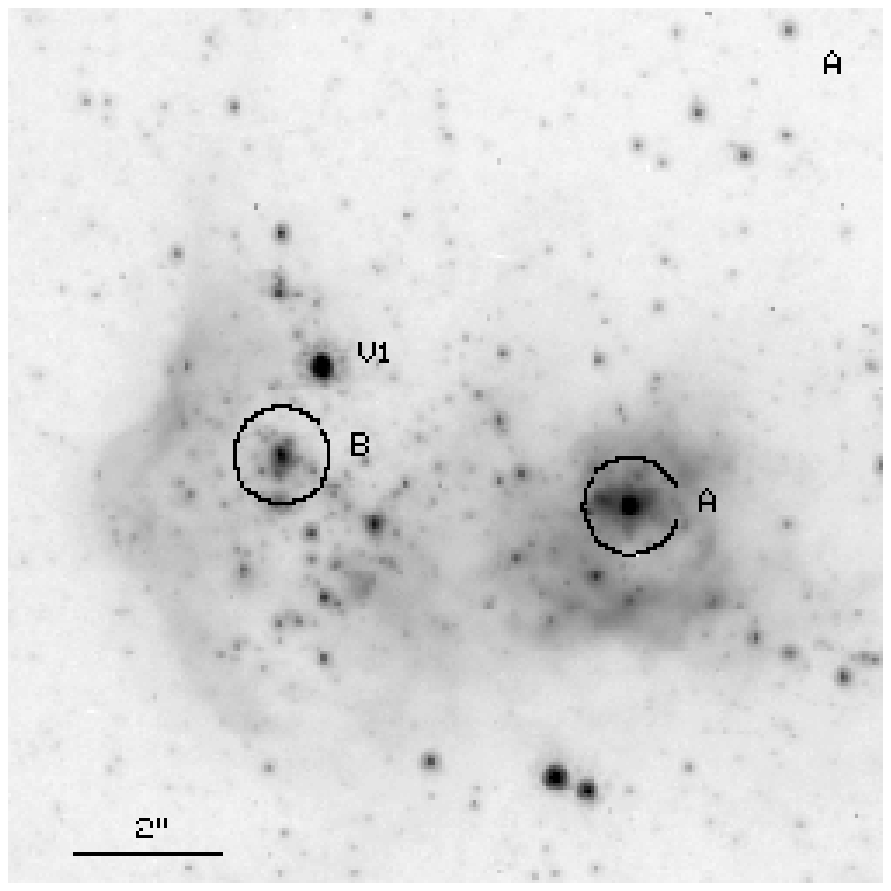}
\plotone{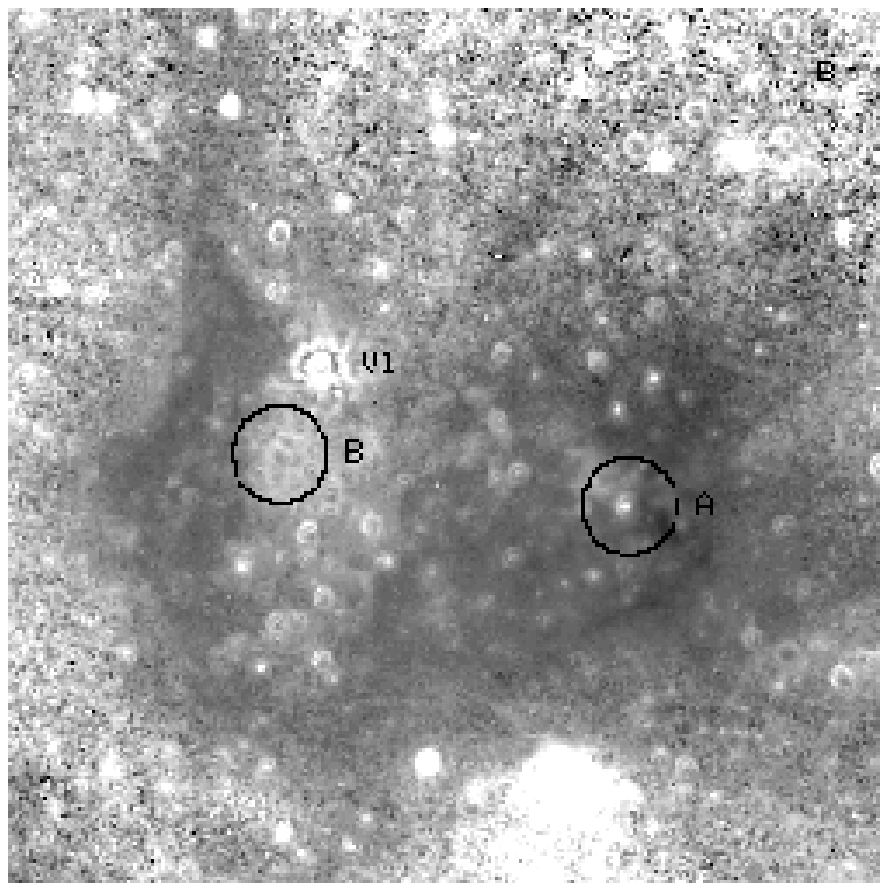}
\caption{PC $I$ (a) and $V-I$ (b) mosaic images of region I (= 
NGC 2363 $\equiv$ Mrk 71) shown with a logarithmic brightness scale. 
North is up, and east is to
the left. In (b) black is blue and white is red. 
Clusters A and B and the luminous blue variable (LBV) star V1 are 
labeled by circles. Note the 
extended red (presumably dusty) region
to the south of cluster A. 
\label{Fig8}}
\end{figure}

   A more detailed view of region I in $I$ and $V-I$ is shown in 
Fig. \ref{Fig8}. Although both clusters A and B are very compact 
(Fig. \ref{Fig8}a), they are marginally resolved.
The FWHM of circular-shaped cluster A on the $I$ image (which is not 
saturated) is 3.8 pixels or 2.8 pc,
similar to the super-star cluster (SSC) R 136a in the Large
Magellanic Cloud and to SSCs in other galaxies. The FWHM of elongated-shaped
cluster B is larger, being $\sim$ 4.7 pixels or $\sim$ 3.5 pc.
Both clusters are blue as evidenced
from the $V-I$ image (Fig. \ref{Fig8}b) with cluster A being slightly redder
because of enhanced dust extinction \citep{Dr00} and saturation of the
central pixels of its $V$ image.
The relatively high internal extinction in cluster A 
($A_V$ $\sim$ 0.3 mag) is confirmed by spectroscopic observations 
\citep{Ma91,Go94,Iz97,Hu99,No00}.
In fact, Fig. \ref{Fig8}b shows that extinction is not confined
to cluster A. 
It is also present in the
extended red region (white in Figure \ref{Fig8}b) to the south
of cluster A. The red color of this region is due partly to two bright
RSG stars and a few other fainter stars, but also to dust. 

We have performed aperture photometry of both clusters
using an aperture with a 6-pixel radius. The background was measured 
within an annulus with radii 6 and
8 pixels and subtracted. Stellar crowding and a highly variable background
limit the precision of the photometry. The derived magnitudes are
corrected for Galactic extinction using $A_V$ = 0.12 mag \citep{S98}
and transformed to the standard $VI$ system using \citet{Ho95b}'
prescriptions. For cluster A we obtain 
$I$ = 17.97 $\pm$ 0.02.
As for cluster B, $V$ = 18.64 $\pm$ 0.02,
$I$ = 18.92 $\pm$ 0.02 and $V-I$ = --0.28 $\pm$ 0.03.
Since our $V$ image of cluster A is saturated, we cannot compare
our photometry with that of \citet{Dr00}. 
However for cluster B our $V$ magnitude is 0.3 mag fainter. 
To check our photometry, we have retrieved
from the {\sl HST} archives the \citet{Dr00} images, and have derived
from them $V$ = 18.66 $\pm$ 0.03 for cluster B, in excellent agreement
with the measurements from our image.
The difference is probably due to a larger aperture used by 
\citet{Dr00} although they do not precise it in their article. 
With a distance modulus $m-M$ = 27.67, the absolute $I$ 
magnitudes of clusters A and B are respectively --9.70 mag and --8.75 mag.

\begin{figure}
\figurenum{9}
\epsscale{1.0}
\plotone{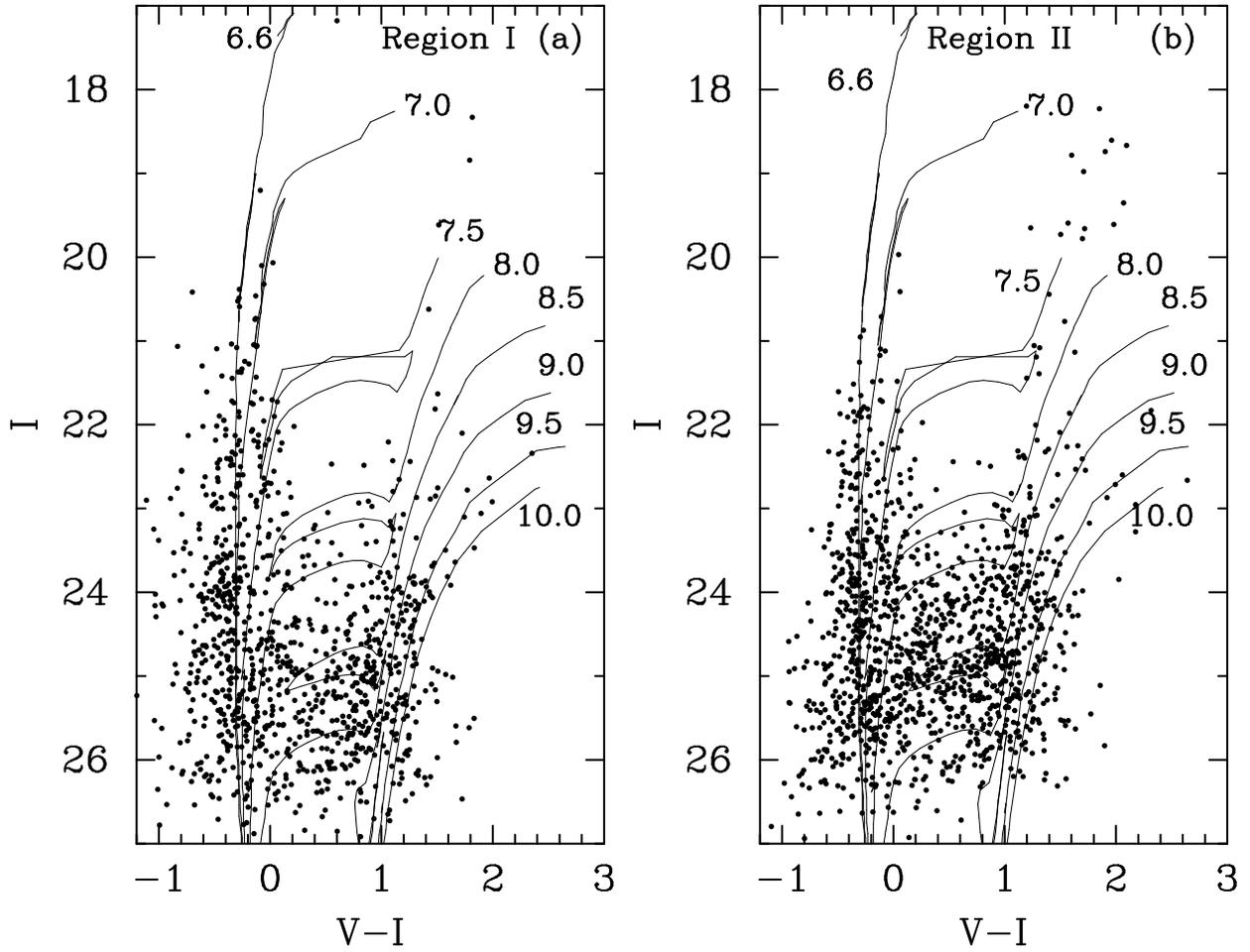}
\caption{$I$ vs $V-I$ CMDs of regions I (a) and II (b) in Fig. \ref{Fig1}. 
Solid lines
are theoretical isochrones from \citet{Be94} for a heavy element
fraction $Z$ = 0.001. They are labeled by the logarithm of the stellar age
in years. 
\label{Fig9}}
\end{figure}

    Fig. \ref{Fig9} shows the CMDs for regions I and II. 
These are dominated
by young stellar populations. They both contain MS
stars with ages $\la$ 10 Myr. Region I also contains
two RSG stars with ages $>$ 10 Myr (the two brightest stars
with $V-I$ $\sim$ 1.8). 
Fig. \ref{Fig9} shows also a substantial population of more evolved
AGB and RGB stars, with ages ranging from 
$\sim$ 100 Myr to $\ga$ 1 Gyr.

The remarkable feature of the CMD of region II
is the presence of numerous RSG stars. \citet{Dr00}
first noted the presence of these stars and estimated their age to be
$\sim$ 10 Myr. Our comparison with the theoretical isochrones of 
\citet{Be94} for the metallicity $Z$ = 0.001 shows that the ages of 
the RSG stars range between 10 and 30 Myr, in agreement with 
\citet{No00}. Thus, although there is presently a strong burst of 
star formation, many of the stars in regions I and II were born over
a long period of star formation which started $\ga$ 1 Gyr
ago. To quantify our statements, we have calculated the number ratios of
different stellar types with $I$ $\leq$ 25 mag. 
For regions I and II we obtain respectively 
MS/RGB = 3.34 and 2.02,
(BL+RSG)/RGB = 2.16 and 2.17, and AGB/RGB = 0.18 and 0.24.
The values of these ratios confirm our statement that 
young stellar populations are dominant in regions I and II.

\subsection{The main body}

\subsubsection{Color-magnitude diagrams}

\begin{figure}
\figurenum{10}
\epsscale{1.0}
\plottwo{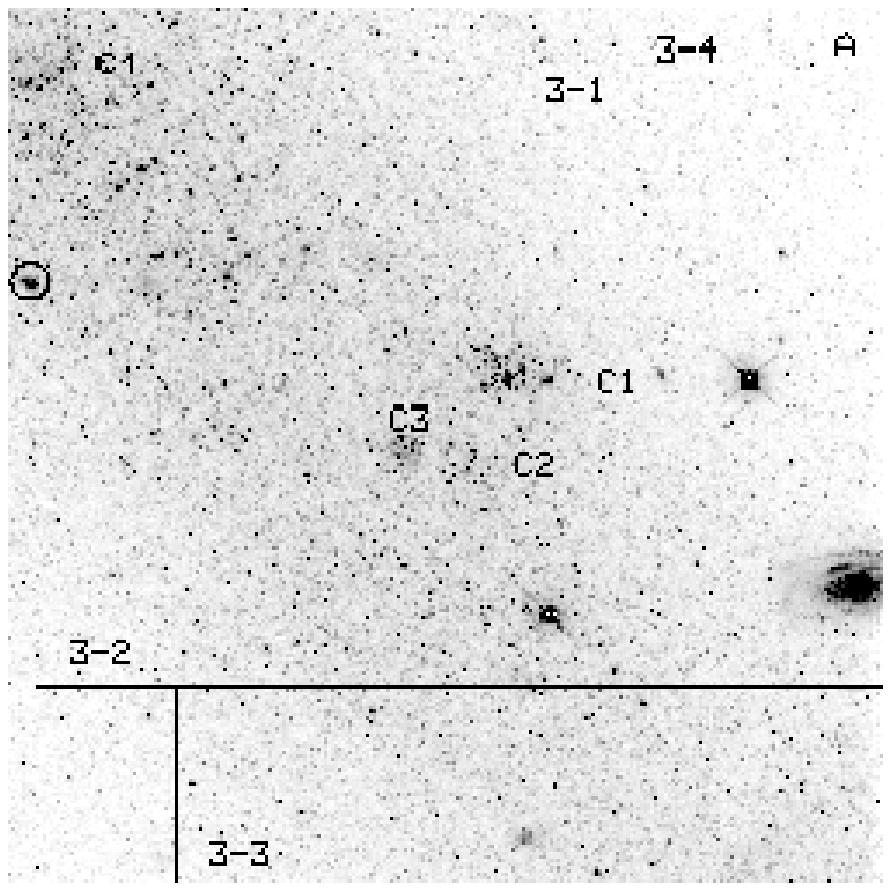}{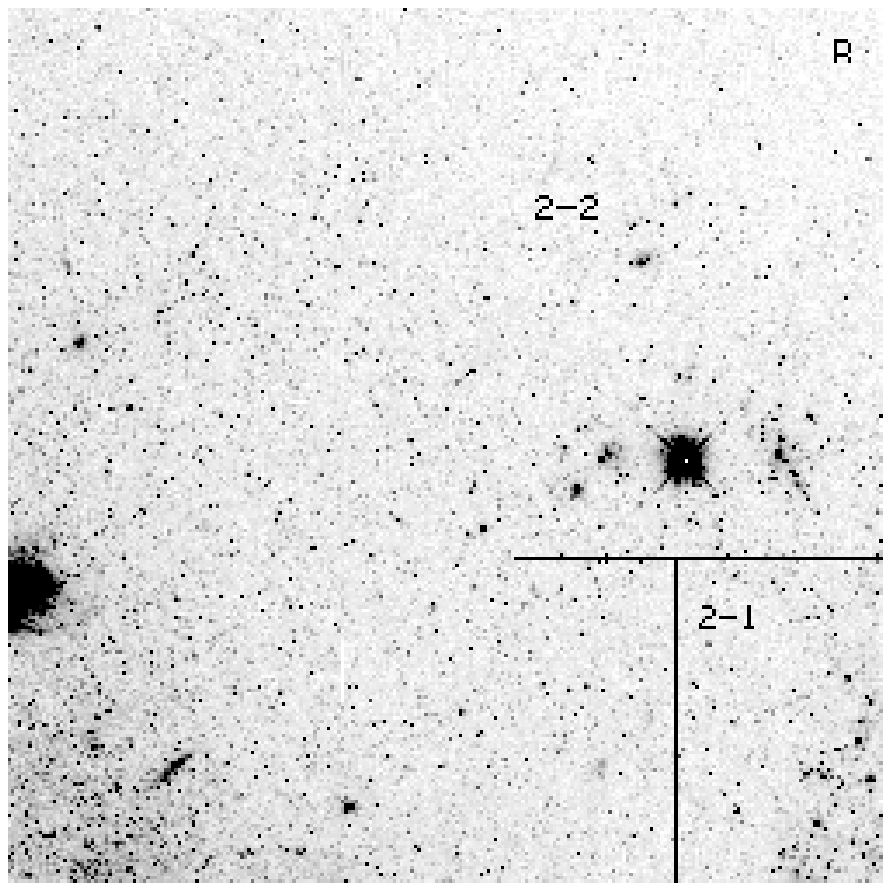}
\caption{Different regions in the WF3 $V$ (a) and WF2 $V$ (b) images 
of NGC 2366
where the CMD analysis of stellar populations has been carried out. 
They are labeled 3-1 to 3-4. Stellar clusters
are labeled C1 to C4 and a compact H {\sc ii} region is enclosed by the circle
in (a). The orientation is the same as in Fig. \ref{Fig1}.
\label{Fig10}}
\end{figure}

   To study variations in the spatial distributions of the stellar populations 
in the main body 
located in the WF3 frame, it is convenient to divide it into three regions 
labeled 3-1 to 3-3,
with the latter being closest to regions I and II discussed 
in the preceding section 
(Fig. \ref{Fig10}a). Region 3-4 which is outside the densest part of
the main body is used as
comparison. The four most prominent stellar open clusters 
are labeled C1 to C4 and the circle indicates a compact H {\sc ii} 
region.
Regions located to the west of the 
main body in the WF2 frame are shown in Fig. \ref{Fig10}b. They
will be discussed later. 

   Region 3-1, which is the furthest away from regions I and II, also contains
the largest number of
MS stars after those regions, as shown by Fig. \ref{Fig6}a.
Several H {\sc ii} regions are present in 3-1 as seen from the 
H$\alpha$ image of \citet{Dr00}. There is
only one high-surface brightness H {\sc ii} region (open circle), probably
ionized by a few O stars. We measure for it 
$V$ = 20.23 $\pm$ 0.02 and 
$I$ = 20.75 $\pm$ 0.05 in an aperture with a 8-pixel radius. The derived color
$V-I$ = --0.52 $\pm$ 0.05 mag of the H {\sc ii} region is bluer than 
that of the  
hottest MS stars, because of a significant contribution of 
[O {\sc iii}] $\lambda$5007 emission in the $V$ band.

\begin{figure}
\figurenum{11}
\epsscale{0.7}
\plotone{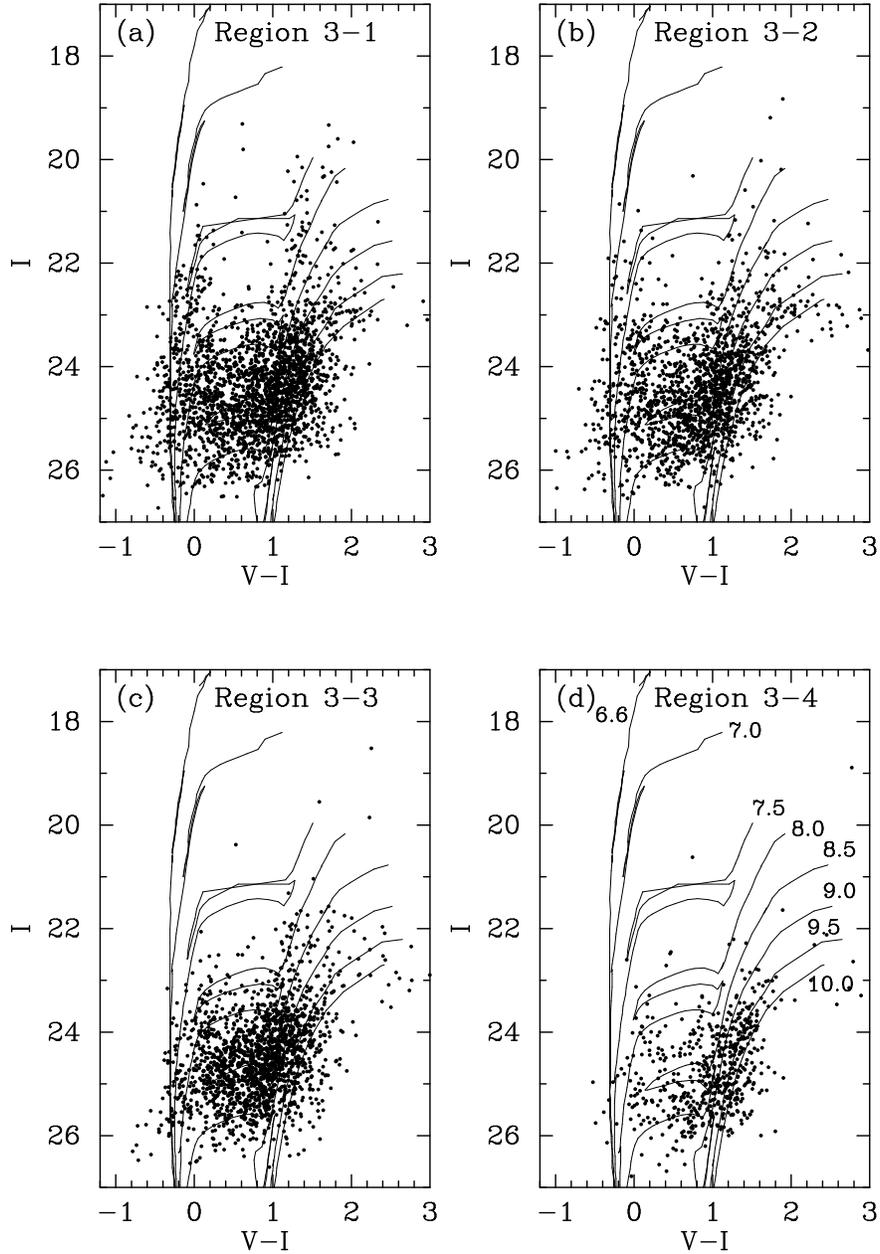}
\caption{$V-I$ vs $I$ CMDs for different regions in the WF3 frame, 
labeled as shown in 
Fig. \ref{Fig10}a and overplotted by theoretical isochrones for a heavy
element mass fraction $Z$ = 0.001 \citep{Be94}. The logarithm of the stellar 
age in years for each isochrone is shown in (d). Stellar populations of various
ages between $\la$ 30 Myr and $\ga$ 1 Gyr are present in regions 3-1 -- 3-3, 
suggesting 
star
formation during the last few Gyr. Region 3-4 is populated 
mainly by a relatively
old stellar population with age $\ga$ 300 Myr.
\label{Fig11}}
\end{figure}

   The number of MS stars in regions 3-2 and 3-3 is
smaller, judging from Fig. \ref{Fig6}a. They are mainly populated
by relatively young BL and RSG stars (Fig. \ref{Fig6}b).
In Fig. \ref{Fig11} we show the CMDs for all regions from 3-1 to 3-4. The main
sequence is most populated in region 3-1 (Fig. \ref{Fig11}a). Other
relatively young stars with ages between 10 Myr and 100 Myr are well
represented as well. Older AGB (age $\ga$ 100 Myr) and RGB (age $\ga$ 1 Gyr) 
stars are also present. The star count ratios 
for this region are  MS/RGB = 0.27, (BL+RSG)/RGB = 1.32 and AGB/RGB = 
0.36. It is seen that, while the
(BL+RSG)/RGB and AGB/RGB number ratios in region 3-1 are comparable to those
in the brightest regions of NGC 2363, regions I and II,  the number ratio 
of MS to RGB stars is about one order of magnitude lower.
The MS in region 3-2 is not as populous as in region 3-1 
with MS/RGB = 0.19
(Fig. \ref{Fig11}b). Only a few post-main-sequence stars with 
ages $\la$ 30 Myr
are seen. The lack of high-surface brightness H {\sc ii} regions in region 3-2
suggests that star formation here has stopped more than 10 Myr ago. However,
the population of BL+RSG and AGB stars is as numerous here as in region 3-1, 
with (BL+RSG)/RGB = 1.08 and AGB/RGB = 0.36,
suggesting that star formation in region 3-2 was active several hundred Myr ago. Only stars with
ages $\ga$ 50 Myr are present in region 3-3 (Fig. \ref{Fig11}c). 
Fewer main-sequence stars are present in this region, giving 
MS/RGB = 0.08.
As in region 3-2, the BL+RSG and AGB 
stellar populations are numerous, so that (BL+RSG)/RGB = 1.12 and 
AGB/RGB = 0.22. Lastly, mainly 
relatively old stars with ages $\ga$ 300 Myr are present in the CMD
of region 3-4 (Fig. \ref{Fig11}d), resulting in small
values of MS/RGB and (BL+RSG)/RGB, 0.03 and 0.50 respectively, and 
a relatively high value of AGB/RGB equal to 0.24. 

\begin{figure}
\figurenum{12}
\epsscale{0.7}
\plotone{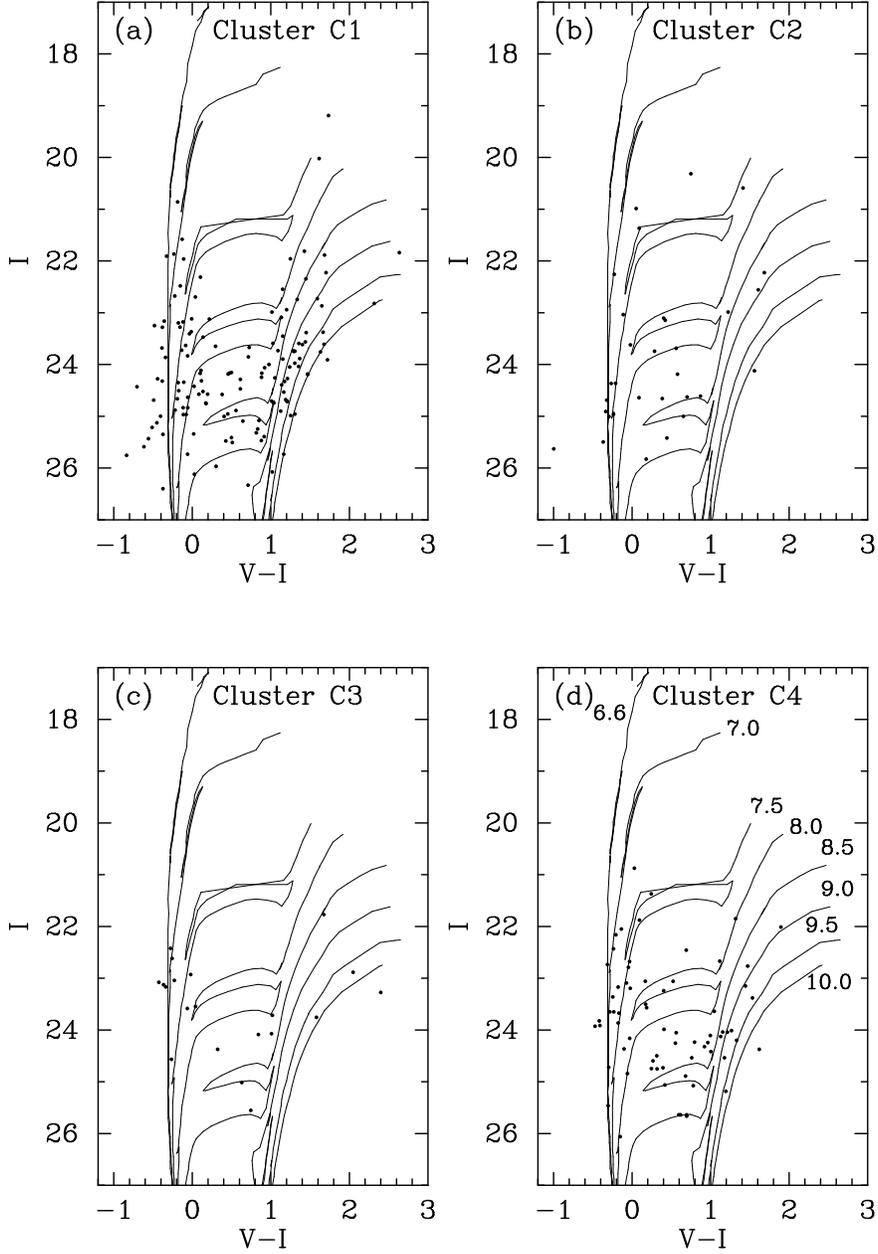}
\caption{$V-I$ vs $I$ CMDs for the stellar clusters labeled in 
Fig. \ref{Fig10}a. They are overplotted by theoretical isochrones for a heavy
element mass fraction $Z$ = 0.001 \citep{Be94}. The logarithm of the stellar
age in years for each isochrone is shown in (d).
\label{Fig12}}
\end{figure}

Thus it is clear that there is a
systematic increase in stellar population age from 3-1 to 3-3,
suggesting propagating star formation in the main body from the SW to the NE
directions, which originated in 3-3 some $\la$ 100 Myr ago. Regions I and II
to the SW of 3-3 do not belong to this age sequence. This is evidence that
new centers of star formation in a cometary-like galaxy like NGC 2366 arise
in the body of the galaxy in a stochastic manner. The numerous
peaks and minima in the neutral H {\sc i} distribution of NGC 2366 
\citep[e.g.][]{T04} are also evidence for this stochastic mode of 
star formation, the present 
star-forming centers coinciding with the H {\sc i} peaks and
the H {\sc i} holes corresponding to cavities carved out in the 
interstellar medium by past starbursts that have faded away.
In Fig. \ref{Fig12} we show the CMDs for the four open
stellar clusters labeled in Fig. \ref{Fig10}a. The absence of 
high-surface-brightness ionized gas emission suggests that the age of all four
clusters is larger than 10 Myr. Several stars with ages between 20 Myr
and 30 Myr are seen in the CMDs of clusters C1, C2 and C3. Cluster C4
appears to be older than 30 Myr.

\subsubsection{Surface brightness profiles}

   Another way to study the properties of stellar populations is to consider
integrated characteristics such as surface brightnesses and colors of 
different regions in the galaxy. The advantage of this approach is that it 
includes both resolved and unresolved stars. The disadvantage is that 
populations with different ages contribute to the integrated light and 
assumptions on the star formation history need to be made to derive the
distribution of stellar ages.

\begin{figure}
\figurenum{13}
\epsscale{1.07}
\plottwo{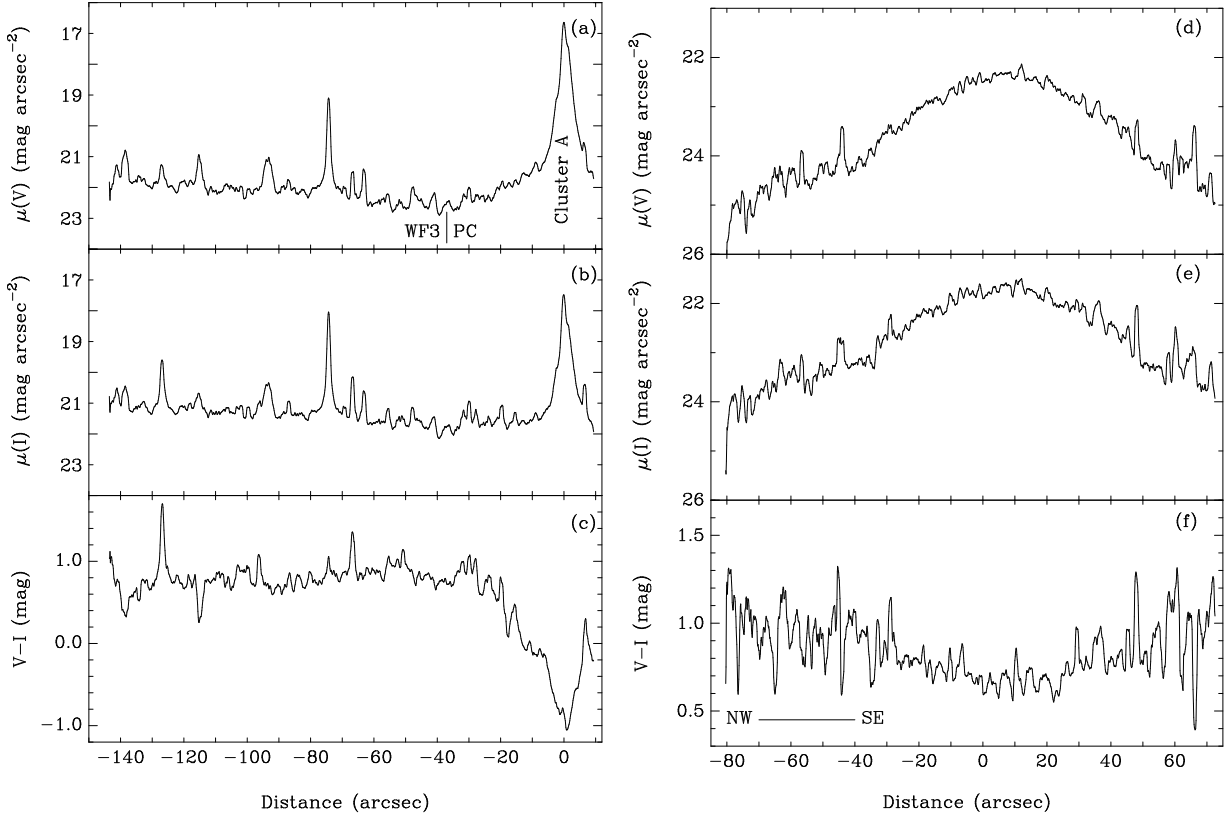}{f13b.ps}
\caption{(a) -- (c) $V$ and $I$ surface brightness and $V-I$ color
profiles along the major axis of NGC 2366 in a
2\arcsec\ wide strip. The origin is set at cluster A, labeled in (a).
The distributions have been smoothed by a 11-point box-car. 
The boundary between the PC and WF3 frames is marked by a vertical line in (a).
The several narrow peaks in the profiles are due to bright stars.
Note the very blue color of the H {\sc ii} region around cluster A (c). The
color along the major axis is relatively constant, with a slight blueing in 
the NE direction.
(d) -- (f) $V$ and $I$ surface brightness and $V-I$ color
profiles perpendicular to the major axis of NGC 2366 in a
10\arcsec\ wide strip and crossing the major axis at the distance of 
--45\arcsec. The origin is set at the intersection of the strips.
The distributions have been smoothed by a 11-point box-car. The perpendicular
color profile shows a clear reddening from the central to the outer parts of
the galaxy (f).
\label{Fig13}}
\end{figure}

   In Fig. \ref{Fig13}a -- \ref{Fig13}c we show the $V$ and $I$ 
surface-brightness
and $V-I$ color distributions along the major axis 
(i.e. along the diagonals of the PC and WF3 frames) of NGC 2366
in a 2\arcsec\ wide strip centered on the star cluster A (Fig. \ref{Fig8})
taken to be the origin. Fig. \ref{Fig13}d -- \ref{Fig13}f show 
the $V$ and $I$ surface-brightness
and $V-I$ color distributions in a 10\arcsec\ wide strip 
perpendicular to the major axis of NGC 2366, with the origin taken to be 
at the intersection of
the strip with the major axis, at the distance of --45\arcsec\ 
as defined in Fig. \ref{Fig13}a -- \ref{Fig13}c.  
Surface brightnesses and colors are transformed to the
standard $VI$ photometric system following the prescriptions
by \citet{Ho95b} and corrected for extinction with $A_V$ = 0.12 mag
\citep{S98}.

It can be seen that cluster A peaks at a surface brightness $\mu$($V$) 
$\sim$ 16.5 mag arcsec$^{-2}$, or $\sim$ 5.5 mag
brighter than the surface brightness in the main body (WF3). 
This peak surface brightness is a lower limit since  
a few central pixels in the $V$ image of cluster A are
saturated. 
The profile of the H {\sc ii} region around A is broader in $V$ 
(Fig. \ref{Fig13}a) than in $I$ (Fig. \ref{Fig13}b) because of the larger
contribution of the extended ionized gas emission in $V$. 
Thus, the region with blue $V-I$ color (Fig. \ref{Fig13}c) is 
particularly broad. The bluest color is $\sim$ --1.0 mag.

   Fig. \ref{Fig13}c shows that there is a slight increase in the $V-I$
color from $\sim$ 0.7 for --140\arcsec\ $\la$ $r$ $\la$ --60\arcsec\ 
(region 3-1) 
to $\sim$ 0.9 for --60\arcsec\ $\la$ $r$ $\la$ --40\arcsec\ (region 3-3). 
We use single stellar population (SSP) models to interpret the colors.
A SSP is composed of stars of the same age formed in an instantaneous 
burst of star formation with masses distributed according to a Salpeter 
initial mass function.  
Instantaneous burst models are more appropriate for fitting 
the integrated colors of the galaxy than continuous star formation models 
with a constant star formation rate
because the star formation rate in different regions of NGC 2366
varied with time. 
This means that, in a given  region, most of the light comes from particular 
stellar types with particular ages.
Thus, in  the
blue regions with $V-I$ $\sim$ 0.7, the surface density of BL and RSG stars is
higher than that of RGB stars (compare Fig. \ref{Fig6}b and \ref{Fig6}d) 
and they contribute most of the light.
The galaxy brightness distribution (Fig. \ref{Fig1}) is similar
to the surface density distribution of BL and RSG stars, but not that 
of the RGB stars. 
On the other hand, in red regions with $V-I$ $\sim$ 0.9, the opposite 
situation prevails:
the surface density of RGB stars is significantly higher than that
of BL and RSG stars and RGB stars contribute most of 
the light of the galaxy in these red regions. 
A $V-I$ color of $\sim$ 0.7 
is consistent with that predicted 
for a $\la$ 3 Gyr single stellar population according
to Padua models with $Z$ = 0.001 \citep[ http://pleiadi.pd.astro.it]{Gi00},
while a $V-I$ color of $\sim$ 0.9 corresponds to a 
single stellar population of several Gyr.
The 
unknown star formation history for NGC 2366 at large ages ($\geq$ 5 Gyr) 
precludes a 
more precise determination of the galaxy's age based on integrated colors.

The minor axis brightness profiles (Fig. \ref{Fig13}d,e) show that the
main body of NGC 2366 has a size of $\sim$ 150\arcsec\ (2.4 kpc)
at a $V$ surface brightness level of 25 mag arcsec$^{-2}$.
This is consistent with the surface photometry of \citet{No00} who found the
size of the optical body to be 5\farcm3 $\times$ 2\arcmin\ at the $B$
surface brightness level of 25 mag arcsec$^{-2}$. H {\sc i} interferometric
mapping reveals that the optical body of NGC 2366 is embedded within a larger
envelope of H {\sc i} gas of size 13\farcm6 $\times$ 4\farcm4 at a H {\sc i}
column density level of 5 $\times$ 10$^{19}$ cm$^{-2}$ \citep{T04}.
With an inclination of $\sim$ 60 degrees as derived from fitting the H {\sc i}
rotation curve by a tilted ring model, this corresponds to a H {\sc i} disk 
of $\sim$ 13.1 kpc in size \citep{T04}. 
As in Fig. \ref{Fig13}c, Fig. \ref{Fig13}f shows an increase 
of the $V-I$ color
from a value of $\sim$ 0.7 in regions near the major axis to a value of 
$\sim$ 0.9 in the outer parts of NGC 2366. This
increase at large distances is
consistent with the more extended spatial distribution of older and hence
redder stellar populations discussed in the introduction to Section 4.

We can also compare the observed $V-I$ color of the reddest regions
in NGC 2366 with the integrated colors of the
globular clusters considered before in the CMD analysis,
the isochrones of which are shown in Fig. \ref{Fig5}. Again, for these regions
which do not contain young stellar populations, instantaneous burst
models constitute a reasonable approximation because of the slow 
variation of the $V-I$ color for ages $\geq$ several Gyr. Also, it is
not excluded that stars in the halo of NGC 2366 
were formed during a relatively
short period. These arguments are supported by the CMD for region 3-4,
where mainly old stellar populations are present. Although this region 
was not used to derive the $V-I$ color spatial
distributions in Fig. \ref{Fig13}, its global color is also $\sim$ 0.9.
From the catalogue of Galactic globular 
clusters of \citet{Ha96}, the $V-I$ colors corrected for extinction
for the clusters M15, NGC 6397, M2, NGC 6752, NGC 1851, 47 Tuc 
listed in order of increasing metallicity are
0.72, 0.80, 0.84, 0.88, 0.98, 1.09 mag. This comparison shows that while a
$V-I$ value of 0.9 mag is still consistent with a 10 Gyr population
with metallicity [Fe/H] = --1.58 (M2) and --1.54 (NGC 6752),
the colors of the more metal-rich clusters NGC 1851 (--1.29) and 
47 Tuc (--0.71)
are too red (see also Fig. \ref{Fig5}a). Therefore, if the metallicity of 
stars in NGC 2366 is larger
than that of NGC 6752, then their age is less than 10 Gyr. 
Such a relatively young age appears to be
typical for cometary galaxies. \citet{No00} derived from ground-based
spectroscopic and photometric observations for several 
cometary galaxies, including NGC 2366, an age of $\la$ 4 Gyr. 

\subsection{Northwestern and southeastern regions}

   We discuss now the CMDs of the northwestern regions located in
the WF2 frame and of the southeastern regions located in the frame WF4.

\begin{figure}
\figurenum{14}
\epsscale{1.0}
\plotone{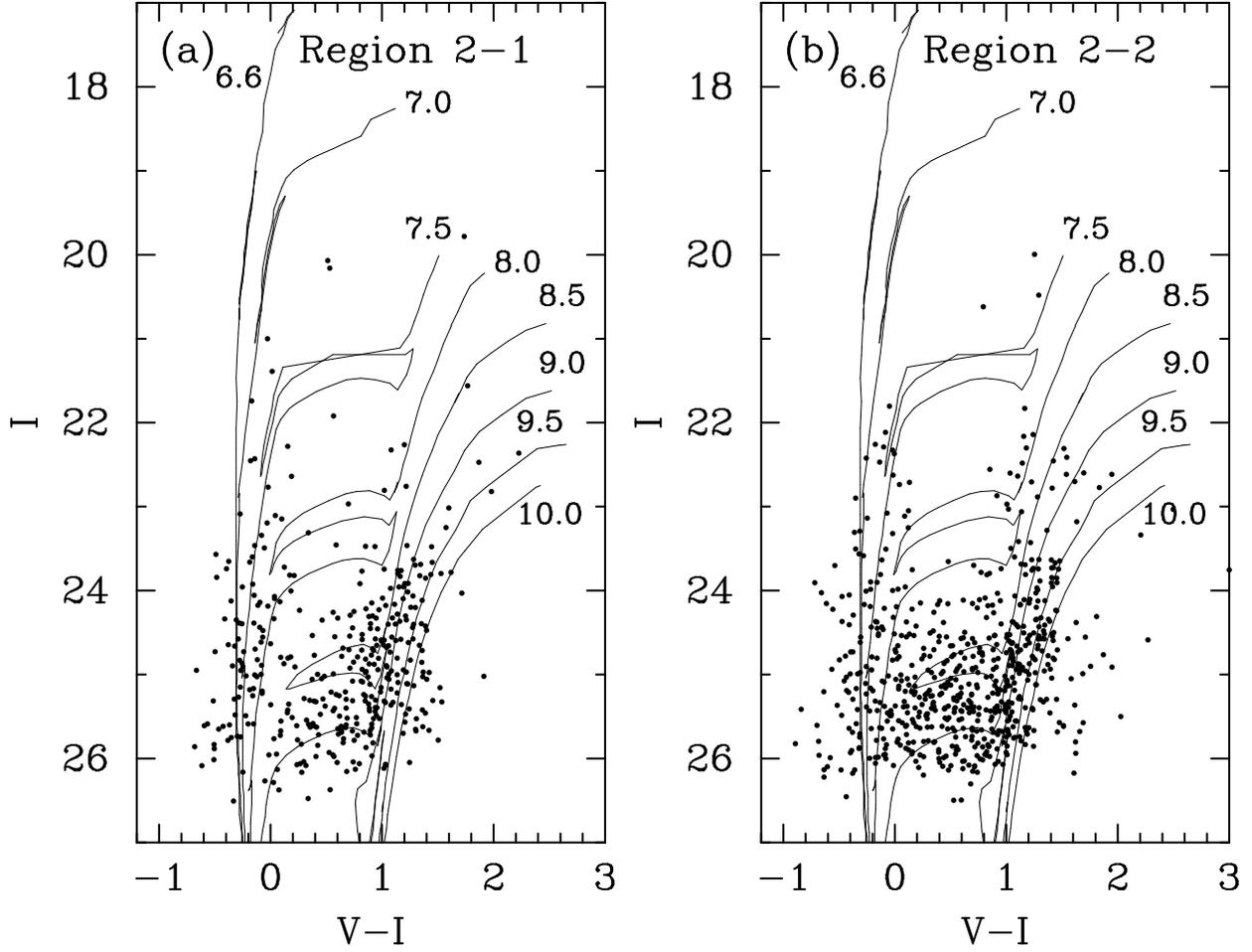}
\caption{$V-I$ vs $I$ CMDs for two regions in the WF2 frame, as labeled in 
Fig. \ref{Fig10}b. They are overplotted by theoretical isochrones for a heavy
element mass fraction $Z$ = 0.001 \citep{Be94}. Each isochrone is labeled
by the logarithm of the stellar age in years.
\label{Fig14}}
\end{figure}

   In Fig. \ref{Fig10}b we show the boundaries of the northwestern region 
2-1 which contains part of region III as defined by \citet{Dr00} and 
of the northwestern region 2-2
where an enhanced density of MS stars is seen 
(Fig. \ref{Fig6}a). The extended low-surface-brightness emission
around some stars suggests the presence of early
B stars. The CMDs of these regions (Fig. \ref{Fig14}) confirm this conclusion.
Several stars with ages in the range 20 -- 30 Myr are present. 
However, AGB and RGB stellar populations are seen as well, implying that
star formation has occurred over the last $\ga$ 1 Gyr in both regions.

\begin{figure}
\figurenum{15}
\epsscale{1.0}
\plotone{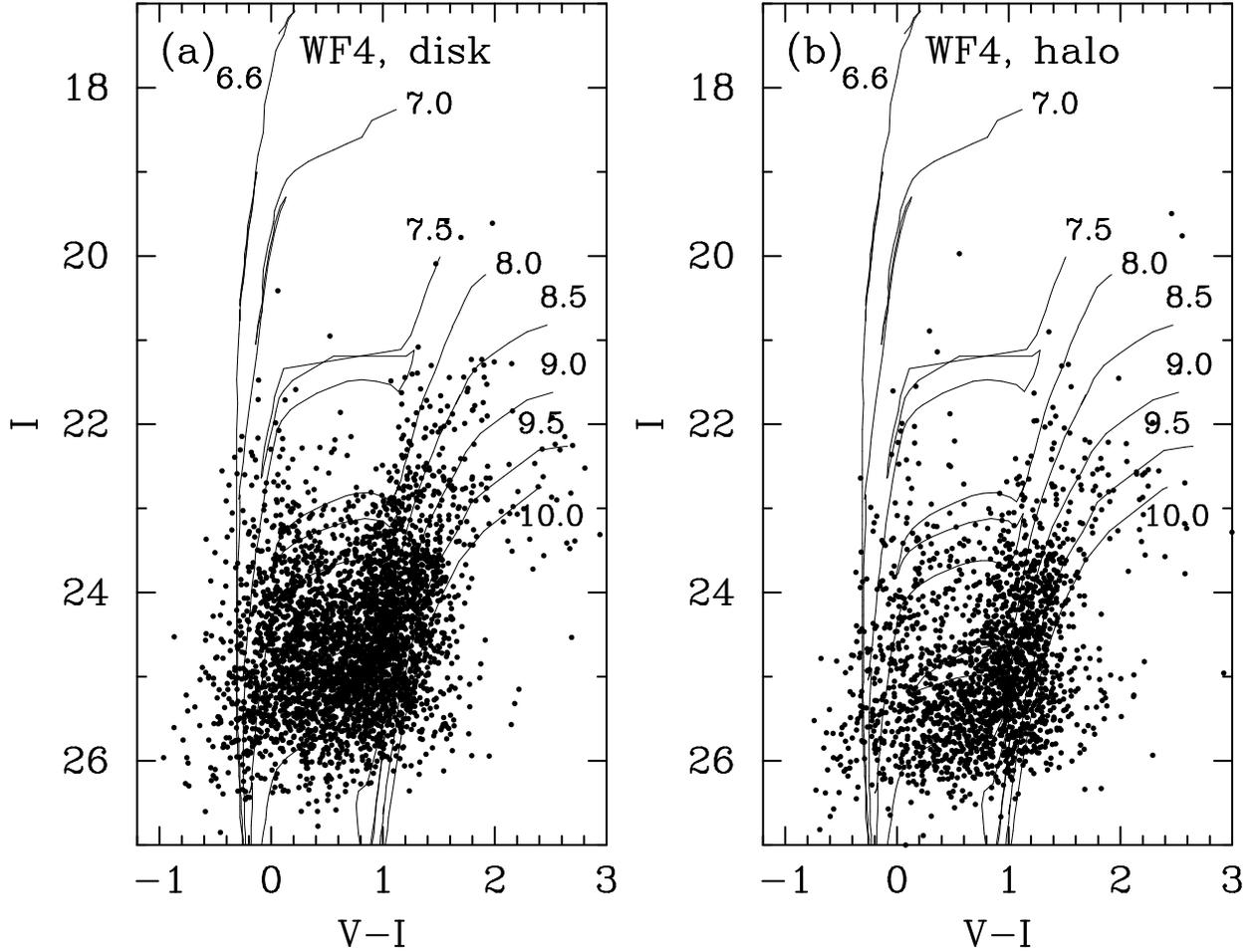}
\caption{$V-I$ vs $I$ color-magnitude diagrams (CMD) 
for the disk and halo regions of NGC 2366 
in the WF4 frame.
The disk CMD includes all stars in 
the region above the diagonal connecting the upper left and 
lower right corners of the WF4 frame in Fig. \ref{Fig1}. The
halo CMD includes all stars in the region below that diagonal.
The CMDs are overplotted by theoretical isochrones for a heavy
element mass fraction $Z$ = 0.001 \citep{Be94}. Each isochrone is labeled
by the logarithm of the stellar age in years.
\label{Fig15}}
\end{figure}

 In Fig. \ref{Fig15} we compare the CMDs of the disk and halo components
of the southeastern region contained in the WF4 frame (Fig. \ref{Fig1}). The 
CMD for the disk component includes sources above the diagonal 
connecting the upper left and lower right corners of the WF4 frame, while
the CMD for the halo component includes sources below that diagonal.
Fig. \ref{Fig15}b shows that the halo is populated mainly by 
old stellar populations with a particularly narrow RGB. On the other hand,
the disk component (Fig. \ref{Fig15}a) is populated by 
intermediate-age stars in addition to older stars.
The fractions of stars of different types which we derived for
the disk and halo are respectively MS/RGB = 0.16 and 0.11, (BL+RSG)/RGB = 1.41
and 0.79, and AGB/RGB = 0.28 and 0.23.

\subsection{AGB and RGB stars}

We now discuss in greater detail the more evolved populations of RGB and AGB 
stars. To put our results in perspective, we will compare the CMD of NGC 2366
with those of other BCDs.

   AGB stars which are tracers of intermediate-age
populations \citep[e.g.,][]{Ma98} are detected in all parts of NGC 2366 
(Fig. \ref{Fig6}). 
There are two striking properties which characterize that AGB stellar 
population. First,  
Fig. \ref{Fig5} shows that they are much more numerous 
in NGC 2366 than in two other BCDs,  
UGC 4483 with a lower ionized gas metallicity (12 + log O/H = 7.54
or 1/23 solar) and at the same distance of 3.4 Mpc \citep{Iz02}, and   
 VII Zw 403 also with a lower ionized gas metallicity 
(12 + log O/H  $\sim$ 7.69 or 1/17 solar)
but at a slightly larger distance \citep[4.5 Mpc as 
determined from the tip of the giant branch,][]{Sc98,Ly98}.
While the ratios MS/RGB and (BL+RSG)/RGB which characterize the young 
stellar populations, vary over a  
wide range from region to region in NGC 2366, the  
AGB/RGB ratio varies in the relatively narrow range of 0.18 -- 0.36,
with an average value of 0.26 for the whole galaxy, counting only stars with 
 $I$ $\leq$ 25 mag. For comparison, the
fraction AGB/RGB is $\la$ 0.14 for VII Zw 403 (this is an upper limit 
because VII Zw 403 is at a larger distance and the RGB star count is less 
complete) and 0.17 in UGC 4483.
The second striking fact
is that AGB stars in NGC 2366 are brighter on the mean than those
in VII Zw 403 and in other dwarf galaxies 
(see Fig. 14 of \citet{Iz02}
where CMDs for five BCDs and irregular galaxies outside the Local Group
and six Local Group irregular galaxies are compared). The AGB stars
in NGC 2366 have a mean absolute $I$ magnitude $\sim$ --4.9 mag
(Fig. \ref{Fig5}), brighter by $\sim$ 0.3 mag than those
in VII Zw 403. However, they are comparable in brightness to those in
UGC 4483 \citep{Iz02}.

The difference in the absolute brightnesses of AGB stars 
in NGC 2366 and VII Zw 403 can be due to 
a younger age in NGC 2366. To derive ages, theoretical 
isochrones are needed. The ones that are available in the literature
fail to
reproduce the observed properties of AGB stars at low metallicities 
in several respects \citep[e.g.,][]{Ly98}. First,
while AGB stars in the CMDs of NGC 2366, of the BCDs VII Zw 403 and UGCA 290
and of some Local Group irregular galaxies have roughly a constant $I$ 
absolute magnitude \citep[see Fig. 14 of][]{Iz02}, 
the models predict a steady increase of absolute $I$ magnitude
with increasing $V-I$ color, up to the tip of the AGB (TAGB) phase when the 
brightness of AGB stars is maximum. The Padua theoretical isochrones 
\citep{Gi00} with 
$Z$ $<$ 0.001 cannot reproduce the red 
colors of the AGB stars seen in these galaxies. As for the Geneva 
isochrones \citep{Le01}, they do not
include the AGB stage. While the Padua isochrones cannot be used to derive 
absolute ages, they are probably good enough for obtaining relative ages. 
We thus compare the AGB populations of NGC 2366
and VII Zw 403, using the Padua models with $Z$ = 0.001 
of \citet{Be94}, and assuming that AGB stars in both galaxies
have the same metallicity. A difference 
of 0.3 mag in brightness translates into an 
age of $\sim$ 3 Gyr for the AGB stars in NGC 2366 assuming an age 
of $\sim$ 10 Gyr for those in VII Zw 403.
\citet{Iz02} have derived an age of $\sim$ 2 Gyr for the AGB stars in
the BCD UGC 4483.

The relative contribution of the bolometric luminosities of
AGB and RGB stars gives information on the star formation history in NGC 2366 
\citep{Re86,Re88,Ma98}.
The method is based on the so-called fuel-consumption theorem \citep{Re86}
which uses as the main ingredient of population synthesis
the amount of nuclear fuel burned in each evolutionary stage. Then 
the bolometric luminosity can be calculated at each evolutionary stage.
The models so calculated predict that, in the case of instantaneous star 
formation, the
total bolometric luminosity of AGB stars is greater than that of RGB stars
if the age of the stellar population is $\la$ 1 -- 2 Gyr \citep{Ma98}. At
ages of $\sim$ 10 Gyr the relative contribution of AGB stars decreases and
becomes 
$\sim$ 4 times lower than that of RGB stars. If instead star
formation has occurred from now to the past, then the 
contribution of AGB stars
is always greater than that of RGB stars \citep{Re86}. 

    Converting the observed $I$ magnitudes of all observed 
AGB and RGB stars to bolometric magnitudes and adding their luminosities, 
we obtain the ratio of
the total bolometric luminosity of AGB stars to RGB stars to be $\sim$ 1.4.
Such a ratio corresponds to an age of $\sim$ 1 Gyr in the case of 
instantaneous star formation, but this age 
is higher if star formation occurs on longer timescales.
It is likely that $\sim$ 1 -- 3 Gyr ago, 
star formation activity in this galaxy was
higher than now. However, we cannot exclude the possibility that 10 Gyr 
old stars are present in NGC 2366. Because of the age-metallicity degeneracy
of the RGB and too large a distance which prevents the 
detection of indicators of old stellar populations, such as RR Lyrae, 
red giant clump and horizontal branch stars, we cannot put a definite upper 
limit to the age of NGC 2366 with our present WFPC2 observations. 
Deeper {\sl HST}/ACS observations are needed. All we can say is that the detected 
stellar populations are consistent
with an age not exceeding $\sim$ 3 Gyr. However, this upper limit is not very
firm because of the uncertain metallicity of AGB and RGB stars.
Finally, we note that the CMD of NGC 2366 which shows an important RGB
stellar population is quite different from the CMD of I Zw 18, the most
metal-deficient BCD known, with a metallicity of only $\sim$ 1/50 that of
the Sun. The CMD of I Zw 18 is conspicuous in its lack of RGB stars, making
the BCD a bona fide young galaxy with an age $\la$ 500 Myr \citep{Iz04}.

\section{Summary}

   We have obtained {\sl Hubble Space Telescope} WFPC2 $V$ and $I$ 
images of the nearby cometary dwarf irregular galaxy NGC 2366 and have studied 
its resolved stellar
population. The analysis of the color-magnitude diagram (CMD) of this galaxy 
has led us to the following conclusions:

   1. The CMD of NGC 2366 is populated by stars with different ages
including young main-sequence stars with ages $\la$ 30 Myr, 
evolved core helium burning massive stars
(blue-loop stars and red supergiants) with ages between
$\sim$ 20 Myr and $\sim$ 100 Myr, and older 
asymptotic giant branch (AGB) and red giant branch (RGB) 
stars with ages $\ga$ 100 Myr and $\ga$ 1 Gyr respectively. 
The most notable feature of the CMD is
the presence of numerous AGB stars. Counting only stars brighter 
than $I$ = 25 mag, we find that, 
while the star count ratios MS/RGB  and (BL+RSG)/RGB which characterize 
the younger stellar populations, vary in a wide range from region to region, 
the ratio AGB/RGB varies over a relatively narrow range of 0.18 -- 0.36,
with an average value of 0.26 for the whole galaxy. For comparison, the
ratio AGB/RGB is $\la$ 0.14 for the BCD VII Zw 403 and 0.17 for the BCD
UGC 4483. This suggests that, in addition to the present
burst of age $\la$ 100 Myr,
strong star formation activity has occurred 
during the last $\la$ 3 Gyr in NGC 2366.

   2. The older the stellar population is, the smoother and more extended 
is its spatial distribution. This is likely due to diffusion and 
relaxation processes of stellar ensembles. An underlying population of 
older stars is found throughout the body of NGC 2366.

   3. From the observed $I$ magnitude of the tip of the RGB equal
to 23.65 $\pm$ 0.10, we derive a distance modulus $m-M$ = 27.67 $\pm$ 0.10
corresponding to a distance 3.42 $\pm$ 0.15 Mpc, in agreement with previous
distance determinations and suggesting that 
NGC 2366 is a member of M81 group of galaxies.


   4. AGB stars in NGC 2366 are appreciably brighter than those observed
in some other dwarf galaxies with similar metallicities. For example, they 
are $\sim$ 0.3 mag brighter than those seen in the BCD 
VII Zw 403. The higher luminosities of the AGB stars in NGC 2366 are probably
due their relatively young ages ($\la$ 3 Gyr) as compared to those in other
dwarf galaxies.

\acknowledgments
The research described in this publication was made possible in part by Award
No. UP1-2551-KV-03 of the U.S. Civilian Research \& Development Foundation 
for the Independent States of the Former Soviet Union (CRDF) and a grant 
No. M/85-2004 of the Ministry of Education and Science of Ukraine.
It has also been supported by NASA grant HST-GO-08769.01-A and
NSF grant AST-02-05785.
Y.I.I. thanks the hospitality of the Astronomy Department of the University of 
Virginia. 





%
\begin{deluxetable}{lrrcrrcrrcrr}
\tablenum{1}
\tablecolumns{12}
\tablewidth{0pt}
\tablecaption{Photometry completeness\tablenotemark{a}
\label{Tab1}}
\tablehead{ Mag. & \multicolumn{2}{c}{PC}&& \multicolumn{2}{c}{WF2}
&& \multicolumn{2}{c}{WF3} && \multicolumn{2}{c}{WF4} \\
\cline{2-3} \cline{5-6} \cline{8-9} \cline{11-12}
 & \multicolumn{1}{c}{F555W} & \multicolumn{1}{c}{F814W} &
& \multicolumn{1}{c}{F555W} & \multicolumn{1}{c}{F814W} &
& \multicolumn{1}{c}{F555W} & \multicolumn{1}{c}{F814W} &
& \multicolumn{1}{c}{F555W} & \multicolumn{1}{c}{F814W} }
\startdata
19--20  & 98.2$\pm$0.9& 98.4$\pm$0.7&
        & 97.9$\pm$0.8& 98.5$\pm$0.7&
        & 97.2$\pm$0.6& 97.3$\pm$0.7&
        & 97.7$\pm$0.6& 98.7$\pm$0.5 \\
20--21  & 97.0$\pm$0.9& 98.1$\pm$0.7&
        & 97.4$\pm$1.1& 98.3$\pm$1.2&
        & 97.3$\pm$0.8& 97.2$\pm$0.9&
        & 97.6$\pm$0.6& 98.0$\pm$1.1 \\
21--22  & 96.7$\pm$0.7& 96.8$\pm$0.6&
        & 97.7$\pm$0.9& 97.7$\pm$0.8&
        & 96.4$\pm$0.8& 96.1$\pm$0.9&
        & 97.5$\pm$1.0& 97.6$\pm$0.4 \\
22--23  & 94.5$\pm$1.0& 94.7$\pm$1.4&
        & 96.8$\pm$0.8& 95.5$\pm$0.9&
        & 94.7$\pm$1.2& 95.6$\pm$1.1&
        & 96.9$\pm$1.0& 95.3$\pm$1.1 \\
23--24  & 91.2$\pm$1.3& 90.6$\pm$1.7&
        & 94.5$\pm$1.2& 92.1$\pm$2.1&
        & 91.5$\pm$1.5& 81.9$\pm$2.1&
        & 93.9$\pm$1.6& 88.6$\pm$1.8 \\
24--25  & 85.5$\pm$1.8& 84.0$\pm$2.6&
        & 90.7$\pm$0.7& 80.2$\pm$2.3&
        & 79.8$\pm$1.2& 58.9$\pm$2.0&
        & 86.7$\pm$1.1& 75.2$\pm$1.9 \\
25--26  & 73.4$\pm$2.5& 65.9$\pm$2.3&
        & 77.6$\pm$1.5& 57.5$\pm$4.6&
        & 50.3$\pm$2.1& 29.9$\pm$2.9&
        & 69.9$\pm$1.8& 48.4$\pm$3.3 \\
26--27  & 59.3$\pm$3.0& 34.3$\pm$3.4&
        & 53.1$\pm$2.2& 14.3$\pm$2.9&
        & 19.4$\pm$2.9&  5.2$\pm$1.9&
        & 43.8$\pm$1.8& 13.0$\pm$1.6 \\
27--28  & 30.2$\pm$3.1&  2.3$\pm$2.6&
        & 23.5$\pm$3.5&  0.2$\pm$0.9&
        &  1.7$\pm$2.1&  0.3$\pm$0.3&
        & 14.2$\pm$2.4&  0.2$\pm$0.3 \\
28--29  &  3.9$\pm$2.7&  0.1$\pm$0.4&
        &  0.8$\pm$0.6& \multicolumn{1}{c}{\nodata}&
        &  0.2$\pm$0.6& \multicolumn{1}{c}{\nodata}&
        &  1.4$\pm$1.4& \multicolumn{1}{c}{\nodata} \\
\enddata
\tablenotetext{a}{Expressed in percentage of recovered stars.}
\end{deluxetable}

\clearpage


\end{document}